\documentclass[12pt,letterpaper]{article}
\usepackage{jheppub}
\usepackage{subfigure}

\title{Light-sheets and AdS/CFT}

\author[a,b]{Raphael Bousso,}
\affiliation[a]{Center for Theoretical Physics and Department of Physics,\\
 University of California, Berkeley, CA 94720, U.S.A.}
\affiliation[b]{Lawrence Berkeley National Laboratory, Berkeley, CA 94720,
  U.S.A.}

\author[c]{Stefan Leichenauer,}
\affiliation[c]{California Institute of Technology, Pasadena, CA 91125, U.S.A.}

\author[a,b]{and Vladimir Rosenhaus}

\abstract{One may ask whether the CFT restricted to a subset $b$ of the AdS boundary has a well-defined dual restricted to a subset $H(b)$ of the bulk geometry.  The Poincar\'e patch is an example, but more general choices of $b$ can be considered.  We propose a geometric construction of $H$.  We argue that $H$ should contain the set $C$ of causal curves with both endpoints on $b$.  Yet $H$ should not reach so far from the boundary that the CFT has insufficient degrees of freedom to describe it.  This can be guaranteed by constructing a superset $L$ of $H$ from light-sheets off boundary slices and invoking the covariant entropy bound in the bulk.  The simplest covariant choice is $L=L^+\cap L^-$, where $L^+$ ($L^-$) is the union of all future-directed (past-directed) light-sheets. We prove that $C=L$, so the holographic domain is completely determined by our assumptions: $H=C=L$.  In situations where local bulk operators can be constructed on $b$, $H$ is closely related to the set of bulk points where this construction remains unambiguous under modifications of the CFT Hamiltonian outside of $b$.  Our construction leads to a covariant geometric RG flow. We comment on the description of black hole interiors and cosmological regions via AdS/CFT.}

\begin{document}
\maketitle

\section{Introduction}

In the study of the Lorentzian AdS/CFT correspondence~\cite{Mal97,Wit98a,BanDoug98,Bal98,HamKab06,Har11}, one may consider the boundary theory defined on a proper subset of the global conformal boundary.  In this case, it is natural to expect that the bulk dual spacetime manifold may be extendible.  That is, the theory on the subset should describe less than the maximally extended global bulk solution. 

\begin{figure}[tbp]
\centering
\subfigure[]{
	\includegraphics[width=2in]{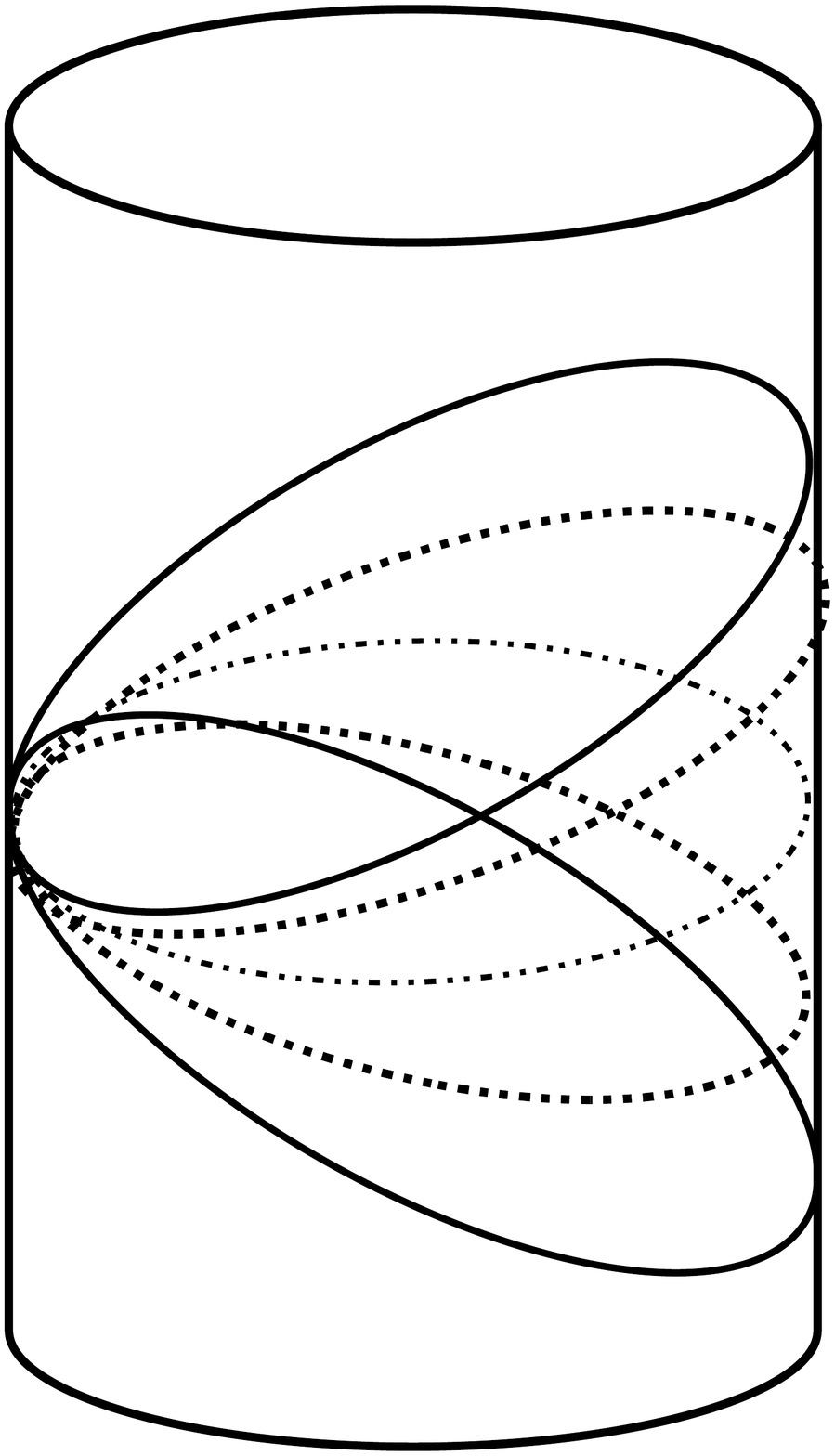}
	}
	\hspace{.5in}
		\subfigure[]{
	\includegraphics[width=2in]{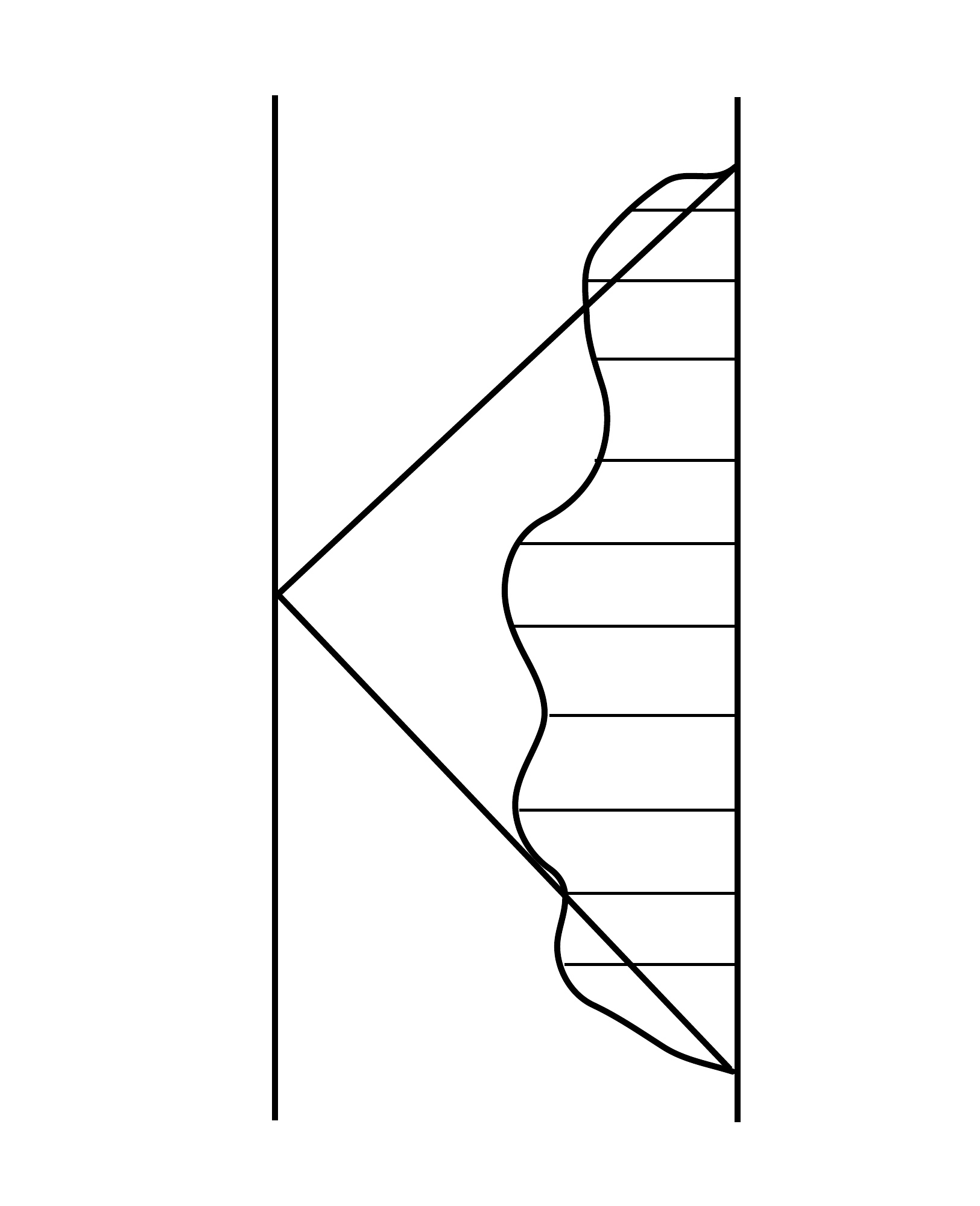}}
\caption{(a) The Poincar\'e patch of AdS, with the usual time slicing in the coordinates of Eq.~(1.1). (b) Time slices of an arbitrary bulk coordinate system that covers the same near-boundary region as the Poincar\'e patch but a different region far from the boundary. This illustrates that there is no preferred coordinate system that would uniquely pick out a region described by the boundary, particularly if the bulk is not in the vacuum state.}
\label{fig-poin}
\end{figure}

The most common example is the CFT on Minkowski space, $\mathbf{R}^{d-1,1}$, which is dual to the Poincar\'e patch of the bulk, with metric
\begin{equation} \label{eq:poin}
ds^2= R^2\frac{-dt^2 + d\vec{x}_{d-1}^2 + dz^2}{z^2}~.
\end{equation} 
These bulk and boundary regions are shown as subsets of global AdS in Fig.~\ref{fig-poin}a.
The Poincar\'e patch seems a ``natural'' choice for the bulk, at least in the above warped-product coordinate system, where the boundary corresponds to $z=0$ and the Poincar\'e patch to $z>0$.  However, there is nothing special about this choice of coordinates.  One could easily write down coordinates that cover a larger or smaller portion of the bulk that is bounded by the same portion of the conformal boundary as $z\to 0$ (see Fig.~\ref{fig-poin}b).  So what selects the Poincar\'e patch as the bulk dual of the CFT on Minkowski space? 

The bulk dual region should be well-defined not only in the vacuum, but for arbitrary states in the semiclassical regime, perturbatively in $1/N$.  Deep in the bulk, the metric will not be that of empty AdS space.  In general all Killing symmetries will be broken, so they cannot be used to pick out a preferred bulk region.  For example, consider a global bulk solution corresponding to a pair of neutron stars orbiting around the origin of the standard global coordinates.  At what time (say, along its worldline) does each star enter and exit the bulk region dual to the CFT on $\mathbf{R}^{d-1,1}$?

Another well-known example is the maximally extended Schwarzschild-AdS black hole. The global conformal boundary consists of two disconnected copies of ${\mathbf R}\times \mathbf{S}^{d-1}$.  In the Hartle-Hawking state (the Euclidean vacuum), the two components can be thought of as slices in a single complex manifold and are related by analyticity.  However, one can consider more general states, for example by adding neutron stars near the left boundary (far from the black hole), and white dwarfs near the right boundary.  Restricting attention to the CFT living on the left boundary, one would expect it to encode the nearby neutron stars, but not the white dwarfs on the far side of the black hole.  But what is the basis of this expectation?  And where does the reach of the left theory end: at the black hole horizon, or somewhere inside the black hole/white hole regions?  Again, there should be an answer to this question that does not rely on special bulk symmetries or coordinate choices.

\begin{figure}[tbh]
\centering
\subfigure[]{\includegraphics[width=1.95in]{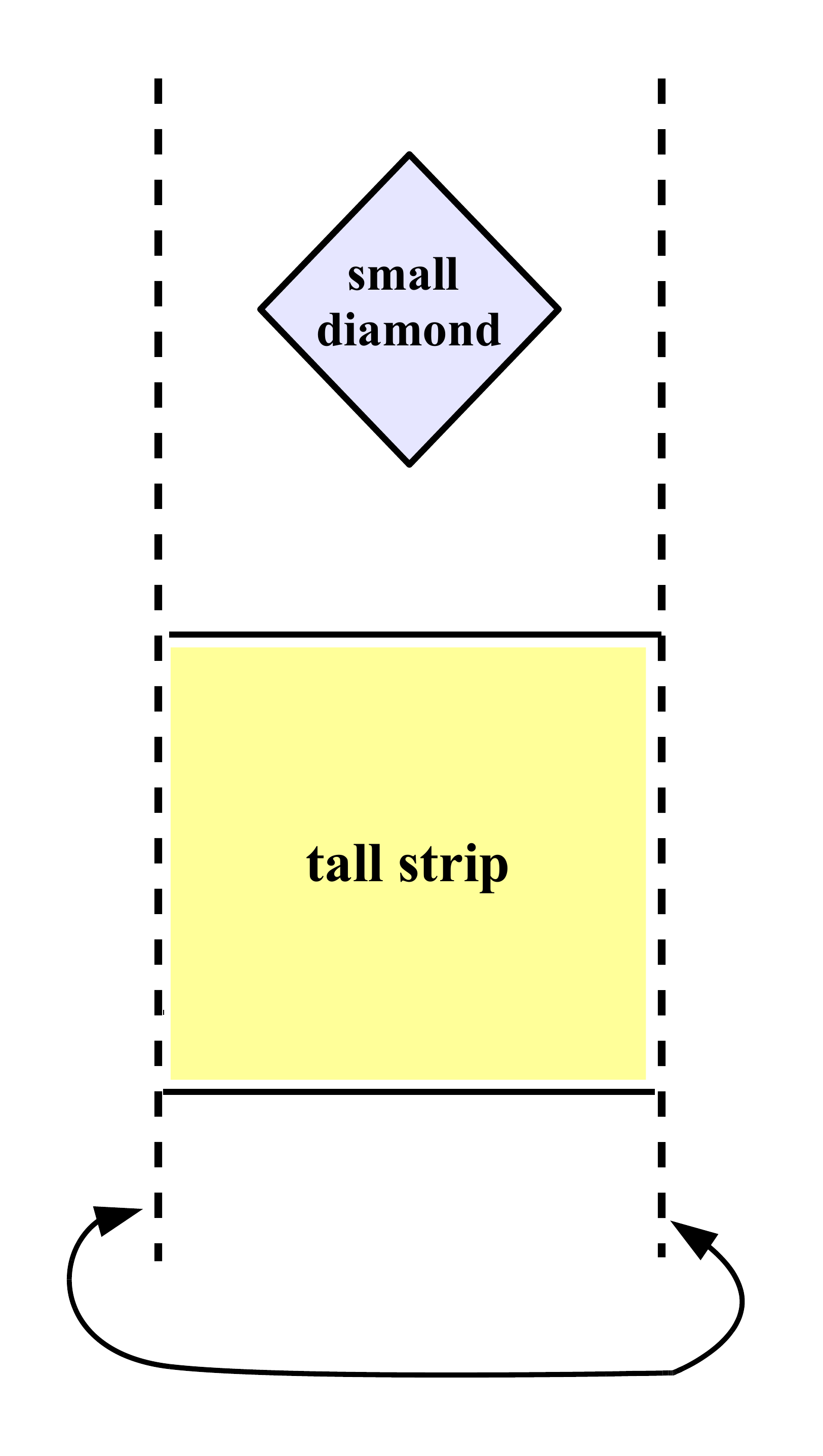}}
\hspace{0in}
\subfigure[]{\includegraphics[width=1.95in]{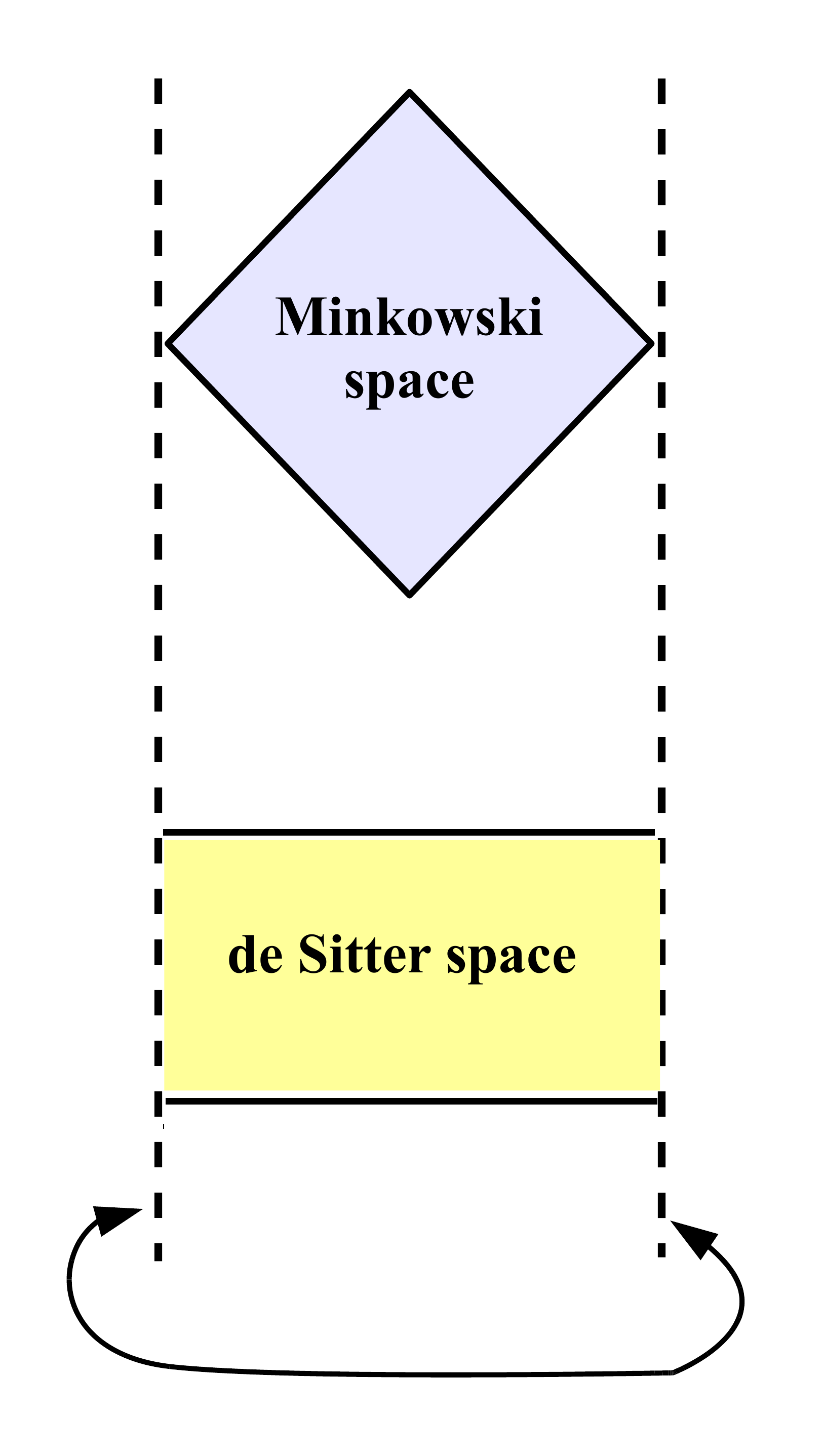}}
\hspace{0in}
\subfigure[]{\includegraphics[width=1.95in]{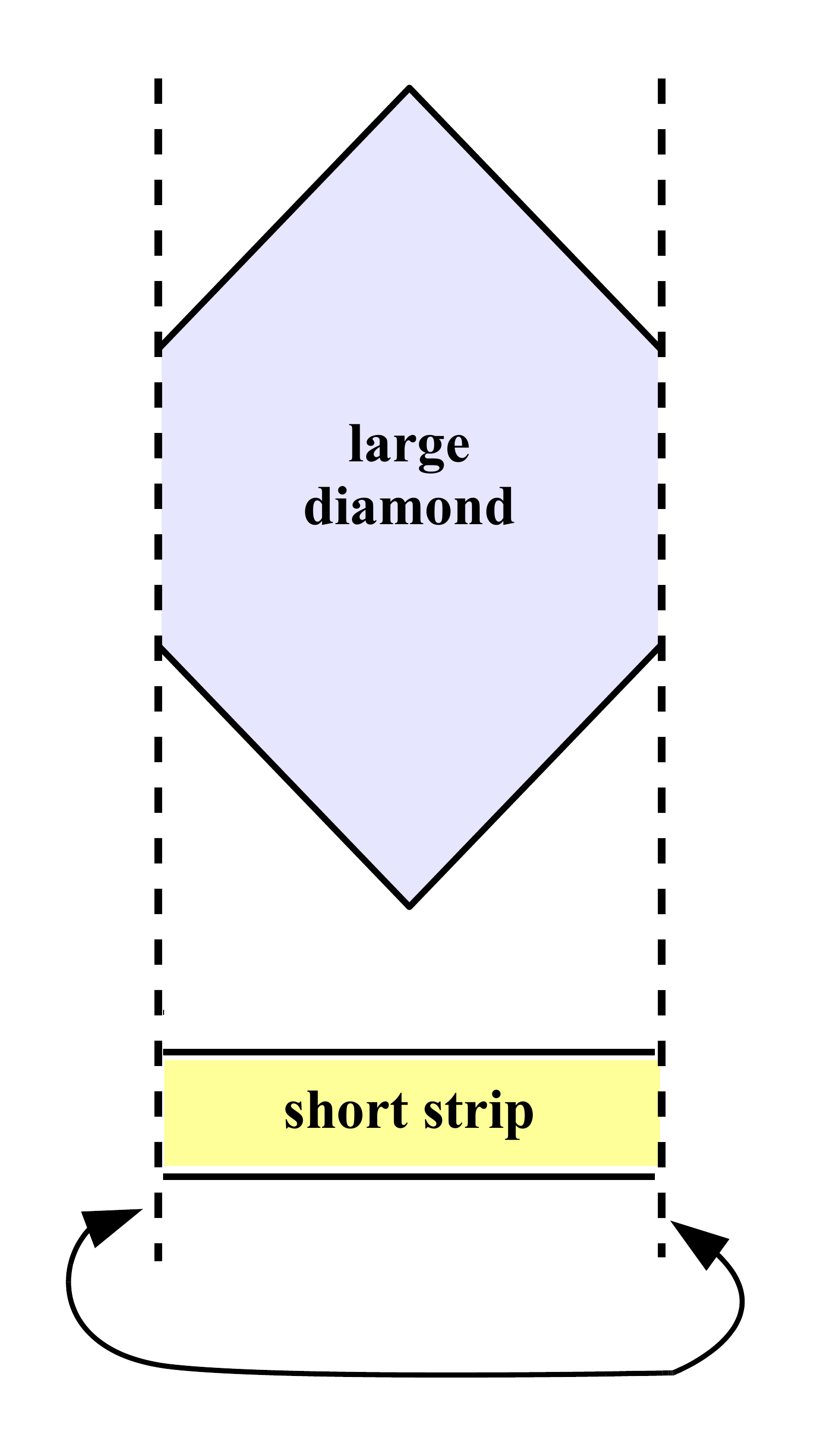}}
\caption{The boundary of AdS; the dashed lines should be identified.  Examples of globally hyperbolic subsets $b$ are shown shaded.  A causal diamond is a set of the form $I^-(q)\cup I^+(p)$, where $q$ is boundary event in the future of the boundary event $p$.  Let $\tau$ be the time along a geodesic from $p$ to $q$ in the Einstein static universe of unit radius ($ds^2=-dt^2+d\Omega_{d-1}^2$).  With $\tau=2\pi$, the causal diamond is the boundary of the Poincar\'e patch.  A causal diamond with $\tau<2\pi$ ($\tau>2\pi$) is called ``small'' (``large'').   An open interval $(t_1,t_2)$ with $t_2-t_1<\pi$ ($t_2-t_1>\pi$ is called ``short strip'' (``tall strip'').}
\label{fig-boundaries}
\end{figure}

There are many other possible choices of subsets $b$ of the global boundary, some of which are shown in Fig.~\ref{fig-boundaries}.\footnote{We consider only subsets $b$ which, viewed as manifolds in their own right, are globally hyperbolic and have the same number of spacetime dimensions as the global boundary.  Otherwise the initial value problem of the CFT would be ill-defined.   If $b$ is a proper subset of the global boundary, then it is not obvious that the CFT on $b$ must have a bulk dual, and we do not set out to prove this or establish under which conditions a bulk dual exists. The question we seek to address is: if $b$ did have a bulk dual region, what would it be? In all cases, a semiclassical bulk dual can be assigned only to some subset of CFT states (excluding, for example, the equal superposition of two global  CFT states corresponding to different bulk metrics), and only perturbatively in $1/N$.  In the case of a small diamond, the bulk dual is the AdS-Rindler patch, so one would expect the geometric states to have a thermal character.}  Let us suppose that the CFT on $b$ describes some portion $H(b)$ of the bulk.  We will call $H(b)$ the holographic domain of $b$.  It should be possible to construct this bulk region geometrically from $b$.  The goal of this paper is to provide such a construction.  
\footnote{In the special case that $b$ is a diamond, motivated by the Ryu- Takayanagi proposal ~\cite{RyuTak06, Tak11, FujTak11} for computing holographic entanglement entropy, one could try constructing $H(b)$ out of minimal surface which start and end on $b$. We will not explore this approach here, however it would be interesting to study how it relates to our construction. We will note however that Ryu - Takayanagi only applies for static situations; for time dependent situations one must use the proposal ~\cite{HubRan07} of (minimal) extremal surfaces, which has been less well established. Additionally, as we will find in Section 6, allowing sufficient modifications of the boundary theory makes the bulk ambiguous in large regions. In this sense, the bulk dual naturally constructed from extremal surfaces would generally be too large. }

\paragraph{Outline}

Our strategy will be to bound the bulk dual region $H(b)$ by a subset $C(b)$ and a superset $L(b)$:
\begin{equation}
C(b)\subset H(b)\subset L(b)~.
\end{equation}
We choose $C(b)$ to be the minimum region the CFT needs to describe to be consistent with bulk causality and locality properties near the boundary.  The set $L(b)$ will be constructed from light-sheets in a way that ensures that the bulk dual  does not have more degrees of freedom (higher maximum entropy) than the CFT.  We will then show that under some additional assumptions on $L(b)$, all three sets are equal.  This fully determines $H(b)$ subject to the stated assumptions.

In Sec.~\ref{sec-causal}, we propose a lower bound on the holographic domain.  We argue that $H(b)\supset C(b)$, where $C(b)=J^+(b)\cap J^-(b)$ is the set of bulk points which lie both in the causal past and in the causal future of the conformal boundary region $b$.  Otherwise, the boundary would fail to describe regions that can be explored by a bulk observer localized near the boundary.  

In Sec.~\ref{sec-holographic}, we propose an upper bound on the holographic domain: $H(b)\subset L(b)$, determined by the requirement that the boundary theory should contain enough degrees of freedom to describe the bulk. The number of CFT degrees of freedom is controlled by the area $A$ of the regulated boundary~\cite{SusWit98}.  In order to ensure that the CFT can describe the holographic domain $H$, $H$ must be contained in a bulk region that satisfies an entropy bound $S\lesssim A$.  In Sec.~\ref{sec-space}, we review the covariant entropy bound, which states that the relevant entropy lies on {\em light-sheets} emanating from the boundary area.  There are no generally valid bounds on the entropy at equal time, so light-sheets should play a preferred role in any holographic duality.  A light-sheet is a null hypersurface generated by nonexpanding light rays orthogonal to a codimension two surface.  

The boundary is of codimension one, so the construction of a bulk region out of light-sheets off the boundary first requires a foliation of the boundary $b$ into time slices (which are codimension two surfaces from the bulk viewpoint).   In Sec.~\ref{sec-light}, we examine whether it is nevertheless possible to define a covariant, slicing-independent upper bound $L(b)$ from light-sheets.\footnote{Light-sheets were used in Ref.~\cite{BouRan01} to determine the holographic domain of (effectively) half of the global boundary.  This region is not globally hyperbolic and the theorem of Sec.~\ref{sec-equiv} does not apply; but the division of the boundary selects for a preferred slicing.  Recently, Hubeny has examined which bulk regions are explored by extremal surfaces of various dimensions in static situations; she finds, as we do for the holographic domain $H$, that black hole interiors are never included~\cite{Hub12}.}

Given a boundary time slicing, one may consider the union $L^+(b)$ of all future-directed light-sheets (one from each slice), or the union $L^-(b)$ of all past-directed light-sheets.  However, both of these sets depend on the slicing of $b$, and the same is true for the union $L^+(b)\cup L^-(b)$.  Thus, neither defines an upper bound on $H(b)$ covariantly. 

We then consider the intersection $L(b)\equiv L^+(b)\cap L^-(b)$ as a candidate for an upper bound on $H(b)$.  There are two apparent problems with this choice.  The first is that $L(b)$, too, would appear to depend on the time slicing of the boundary set $b$.  Secondly, it is easy to see that $L(b)\subset C(b)$.  This would conflict with our expectation that $C(b)\subset H(b) \subset L(b)$, unless it can be shown that $L(b)=C(b)$.  That is, the consistency of our arguments requires that any point that lies on a causal curve that begins and ends on the boundary region $b$ must also lie on both a past and a future-directed light-sheet emanating from $b$, independently of the time slicing of $b$.  

We prove this nontrivial result in Sec.~\ref{sec-equiv}.  As a corollary, the slicing independence of $C(b)$ implies the slicing independence of $L(b)$.  Assuming this is the correct choice of $L(b)$, it follows that the holographic domain $H(b)$ is completely determined:
\begin{equation}
H(b)=C(b)=L(b)~.
\end{equation}

The description of $H(b)$ in terms of light-sheets allows us to define a holographic bulk RG flow, and the equivalence with $C(b)$ makes the RG flow manifestly covariant.  Combining the focussing theorem with the covariant entropy bound guarantees that the number of degrees of freedom is strictly nonincreasing along the flow. We discuss this construction in Sec.~\ref{sec-RG}.

In Sec.~\ref{sec-nonlocal} we consider natural definitions of the bulk dual region from the perspective of the field theory, which contains nonlocal operators that can probe deeply into the bulk.  This approach is less general than our geometric approach, because the construction of such operators is not known for arbitrary $b$.  In cases where it is known we find that a region $\bar H$, very similar to $H(b)$, is picked out as the region with an unambiguous bulk interpretation.

Sec.~\ref{sec-bh} is largely independent of the rest of this paper.  We consider the extent to which AdS/CFT adds to our understanding of quantum gravity in regions with dominant self-gravity, such as black hole interiors and cosmological regions.  We describe a thought-experiment involving the formation and evaporation of (smaller) black holes in such regions, and we argue that its description requires a nonperturbative bulk theory.

\section{The Causally Connected Region $C$}\label{sec-causal}

In this section, we argue that the CFT on the boundary portion $b$ must at least describe the set $C(b)$ of bulk points that are both in the past and in the future of $b$: $H(b)\supset C(b)$.

Consider a bulk excitation very close to the boundary, e.g. at $\frac{\pi}{2} - \rho = \epsilon \ll 1$ in global coordinates,
\begin{equation}
ds^2= \frac{R^2}{\cos^2 \rho}\left(-d\tau^2 + d\rho^2 + \sin^2\rho d\Omega_{d-1}^2\right)~.\label{eq:AdSGlobal}
\end{equation}
Such an excitation is represented on the boundary by excitations with support on a region of size $\epsilon$, at the same transverse position~\cite{SusWit98}.  Here we take the boundary theory to live on a unit sphere, but this property is essentially local.  It holds in Poincar\'e as well as global coordinates.  One expects, therefore, that it should hold for any boundary region $b$, as long as $\epsilon$ is much smaller than the characteristic temporal and spatial size of $b$.  This implies that the boundary theory on $b$ must describe at least the state and the dynamics of a bulk region sufficiently close to $b$. We will exploit the fact that this near-boundary region, in turn, is causally connected to the larger bulk region $C(b)$ to show that the CFT must at least describe $C(b)$.

Consider now a family of bulk observers localized at $\frac{\pi}{2} - \rho = \epsilon \ll 1$.
They will require a finite proper acceleration of order $R^{-1}$, where $R$ is the AdS curvature radius.  The proper time for which this acceleration must be maintained is
\begin{equation}
\Delta t_{\rm proper} = \frac{R}{\epsilon}\Delta \tau~.
\end{equation}
This diverges in the limit as $\epsilon \to 0$, but we will only need to consider the case of small but finite $\epsilon$.  Thus there appears to be no obstruction, in principle, to the existence of such bulk observers. What region can they explore?  We will treat the observers as a collective, imagining that they densely fill Cauchy surfaces of the boundary $b$, moved into the bulk by $\epsilon$.  Such observers can receive signals from the causal past of $b$ in the bulk, and they can send signals to the causal future of $b$.  However, they cannot determine the state in the entirety of either of these regions without making additional assumptions.  

The only region that can be actively explored and manipulated by observers near the boundary is the intersection of the past and the future of $b$,
\begin{equation}
C(b) \equiv J^+(b)\cap J^-(b)~.
\end{equation}
This region can be probed by preparing local probes at an early time, allowing them to travel deeper into the bulk and then back to the observer at a late time.  The outcome of such an experiment is completely determined by the state in $C(b)$, by causality.  And conversely, local fields at any point in $C(b)$ can be manipulated by such an experiment.  Since the experiments of these near-boundary observers are described by the CFT, then for a consistent duality to hold the CFT must describe at least the region $C(b)$:
\begin{equation}
C(b)\subset H(b)~.
\end{equation}
If the near boundary region can both probe and be affected by $C(b)$, then the same must be true for the boundary theory itself.

By contrast, knowledge of the state in $J^-(b)-J^+(b)$, say, is not necessary in order to compute the dynamics near the boundary or in any region that can be explored from the boundary.  It is sufficient to specify initial conditions on the past boundary of  $C(b)$ in the bulk.  Conversely, since its past boundary need not be a Cauchy surface for $J^-(b)-J^+(b)$, the state in $J^-(b)-J^+(b)$ is not determined by the state in the region $C(b)$ which can be explored from the boundary.
\begin{figure}[tbp]
 \centering
 \includegraphics[width=7cm]{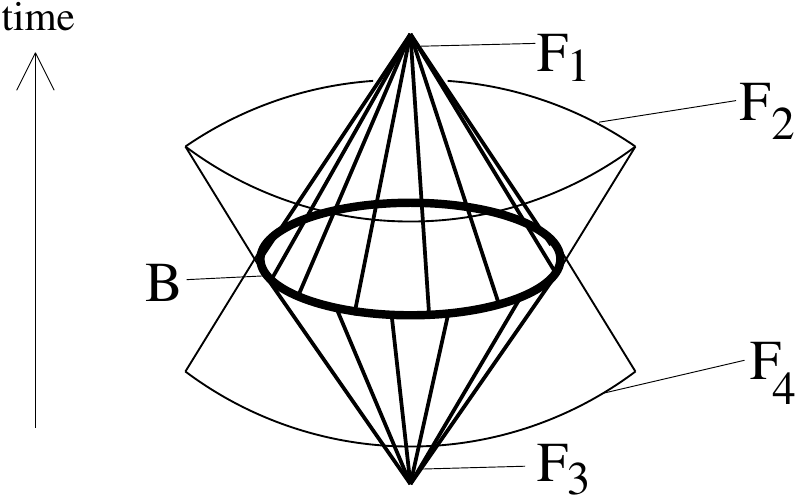}
  \caption{The four null hypersurfaces orthogonal to a spherical surface $B$ in Minkowski space.  The two cones $F_1$, $F_3$ have negative expansion and hence correspond to light-sheets.  The covariant entropy bound states that the entropy of the matter on each light-sheet will not exceed the area of $B$.  The other two families of light rays, $F_2$ and $F_4$, generate the skirts drawn in thin outline.  Their cross-sectional area is increasing, so they are not light-sheets, and the entropy of matter on them is unrelated to the area of $B$.}
\label{fig-spheresheets}
\end{figure}

\section{The Light-sheet Region $L$}
\label{sec-holographic}

In this section we construct a region $L(b)$ which is guaranteed to be encodable in the CFT on the boundary portion $b$, in the sense that holographic entropy bounds guarantee that the maximum entropy of matter in $L(b)$ does not exceed the number of degrees of freedom of the CFT on $b$.  For an exact duality to hold, $H(b)$ must be contained in some region with this property, and we will conjecture that $H(b)\subset L(b)$. 

The holographic principle~\cite{Tho93,Sus95,CEB1,CEB2} is a universal relation between area and quantum information.  This relation manifests itself empirically as the Covariant Entropy Bound~\cite{CEB1}: Let $A$ be the area of an arbitrary (open or closed) codimension-two surface $B$, and let $S$ be the entropy of the matter on any light-sheet $\ell$ of $B$:

\begin{equation}
S[\ell(B)]\leq \frac{A}{4}~.
\label{eq-ceb}
\end{equation}

At the core of the covariant entropy bound is the notion of a light-sheet (Fig.~\ref{fig-spheresheets}).  Light-sheets are null hypersurfaces generated by nonexpanding light rays orthogonal to the surface $B$.  There are four orthogonal directions at every point, since the surface has two sides and we can consider past and future directed light rays.  If the null energy condition is satisfied, at least two of these directions will have nonpositive expansion and thus give rise to light-sheets.  The covariant entropy bound holds separately on each light-sheet.  (For a review, see Ref.~\cite{RMP}.)

\subsection{Spacelike Holography vs Light-Sheets}
\label{sec-space}

The holographic principle is {\em not\/} the statement that the entropy in any spatial region $V$ is limited by the area of the surface bounding that region.  This ``spacelike'' entropy bound follows from the covariant bound in certain special cases~\cite{CEB1}, but in general it is false.  Counterexamples are easily found in cosmology, inside black holes, and even in weakly gravitating systems~\cite{RMP}.  As a general statement valid in all spacetimes, one must not think of holography as effectively projecting out a spacelike direction.  Holography projects along a null direction, just as it does in a conventional hologram.

\begin{figure}[tbp]
\centering
 \includegraphics[width=8.5cm]{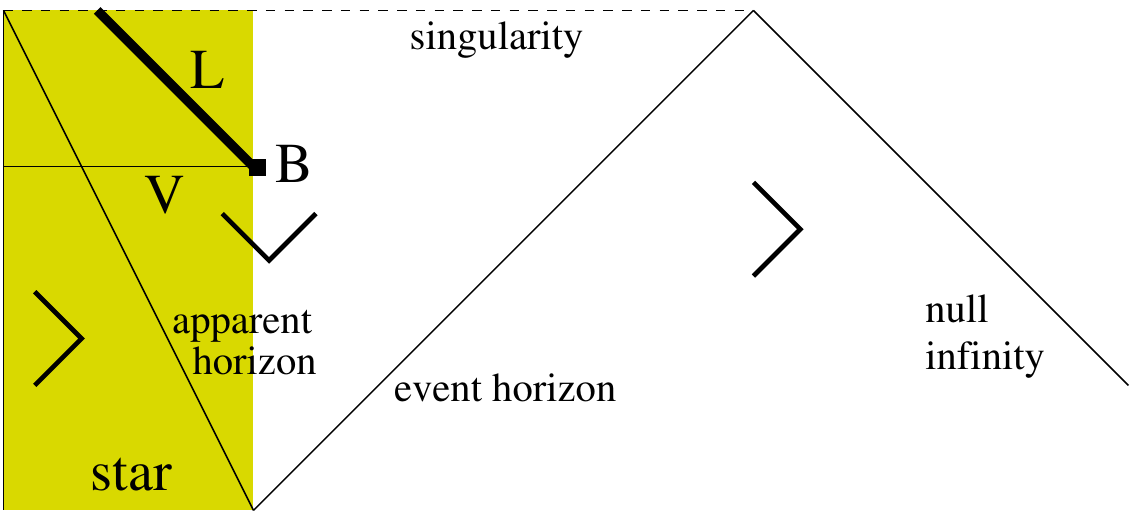} \caption{Penrose diagram of a collapsing star (shaded).  At late times, the area of the star's surface becomes very small ($B$).  The enclosed entropy in the {\em spatial\/} region $V$ stays finite, so that the spacelike entropy bound is violated.  The covariant entropy bound avoids this difficulty because only future directed light-sheets are allowed by the nonexpansion condition. $L$ is truncated by the future singularity; it does not contain the entire star.}
\label{fig-star}
\end{figure}
To illustrate the falsehood of any spacelike notion of holography, consider the surface area of a collapsing star (Fig.~\ref{fig-star}).  The area approaches zero near the singularity, but the entropy of the star starts out finite and cannot decrease.  Thus, $S(V)\gg A$ at late times: a violation of the spacelike entropy bound. However, a light-sheet off of a late-time surface will not penetrate the whole star, so the covariant bound, Eq.~(\ref{eq-ceb}), is upheld.

\begin{figure}[tbp]
\centering
\includegraphics[width=3in]{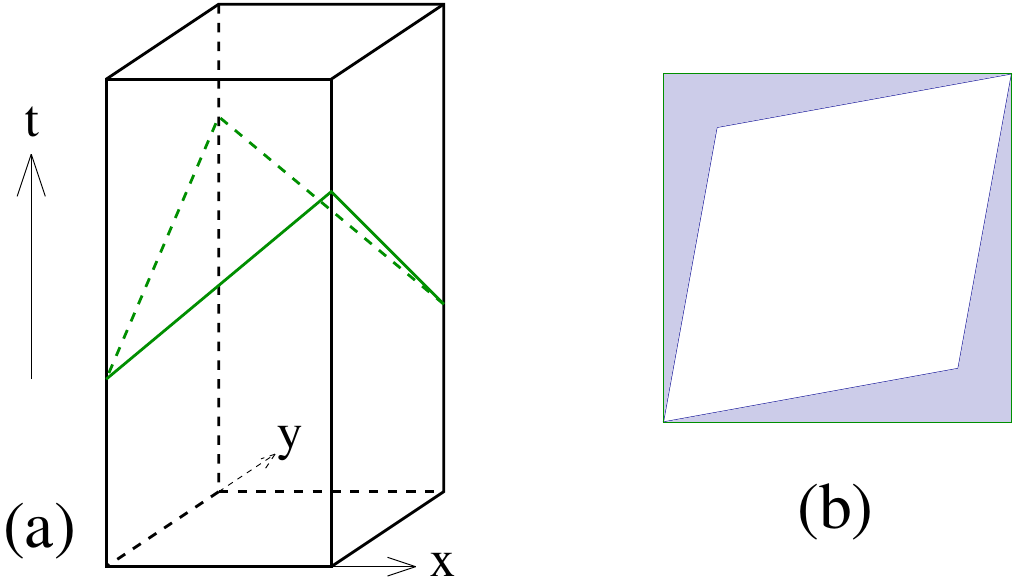}
\caption{(a) A square system in 2+1 dimensions, surrounded by a
surface $B$ of almost vanishing length $A$.  The entropy in the enclosed {\em spatial\/} volume can exceed $A$.  (b) [Here the time
dimension is projected out.] The light-sheet of $B$ intersects only
with a negligible (shaded) fraction of the system, so the covariant entropy bound is satisfied.}
\label{fig-wiggly}
\end{figure}

Another example is shown in Fig.~\ref{fig-wiggly}: It is possible to surround any matter system with a surface of arbitrarily small area.  (For weakly gravitating systems the surface will consist of elements that are highly boosted with respect to the rest frame.)  Thus, the spacelike entropy bound is violated.  By contrast, the light-sheets off of such surfaces do not contain most of the system, evading violation of the covariant entropy bound.  Details and further examples are given in Ref.~\cite{RMP}.  

Notice that the above counterexamples to a spacelike entropy bound can easily be embedded into an asymptotically AdS spacetime.  In particular, consider a timelike hypersurface that is a direct product,  $b={\mathbf R}\times {\mathbf S}^2$.  (For definiteness, we consider $AdS_4$ but our arguments hold in any number of dimensions.)  In the limit where the spatial two-spheres are large, $b$ encloses a very large spacetime region.  We can think of $b$ as a regulated version of the conformal boundary.  The bulk spacetime it encloses is described by a conformal field theory living on $b$, with a UV cutoff on a length scale comparable to the AdS curvature radius~\cite{SusWit98}.  

Yet, the hypersurface $b$ can also be foliated into two-dimensional slices which have arbitrarily small area; and each such slice bounds a global bulk slice.   With this slicing, a naive ``spacelike'' interpretation of holography would seem to imply that the bulk can be described by a theory with an arbitrarily small Hilbert space.  This interpretation is clearly incorrect, as the bulk can have arbitrarily large entropy in the limit where $b$ approaches the boundary.

\begin{figure}[tbp]
\centering
\includegraphics[width=3in]{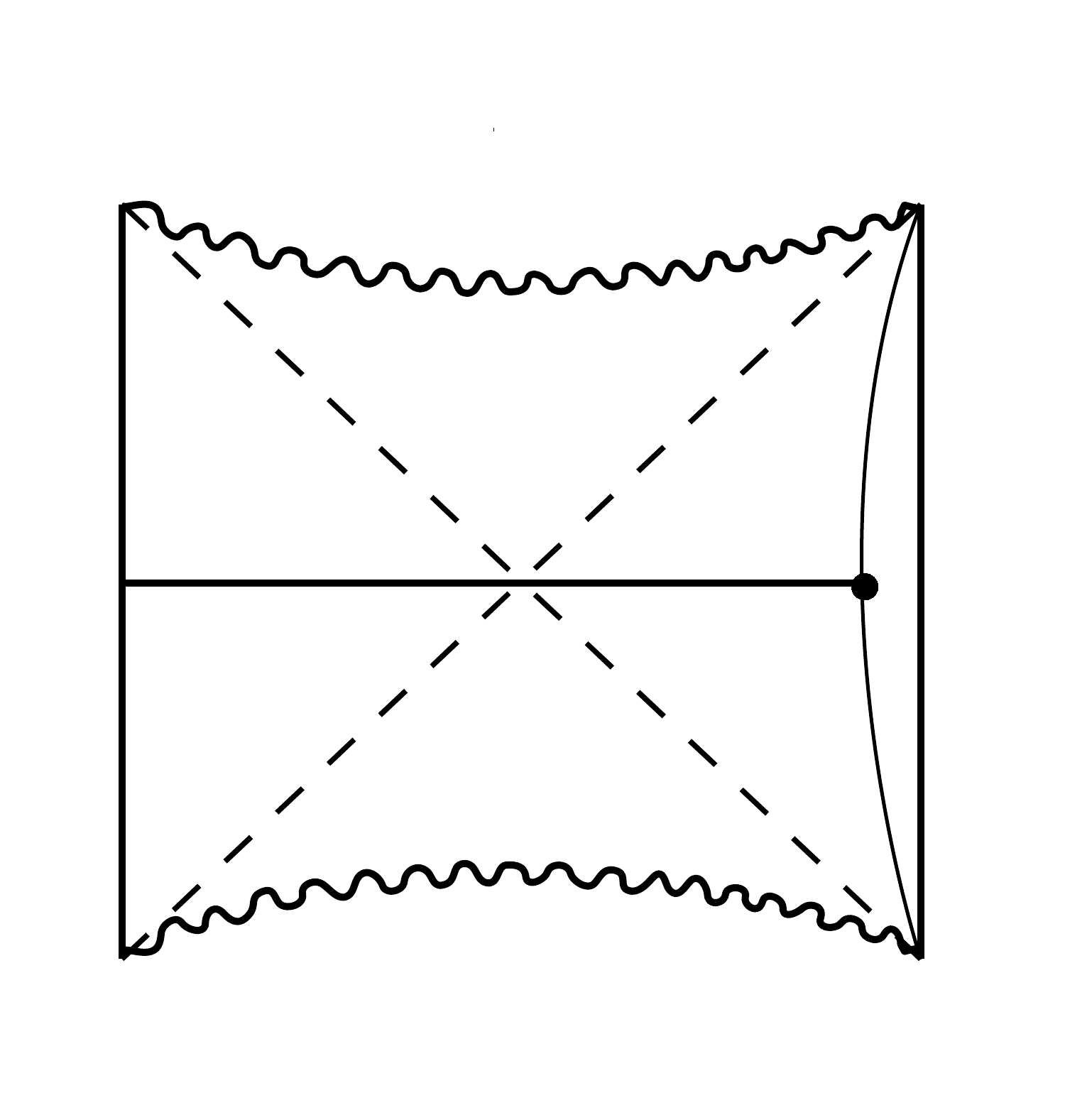}
\caption{An AdS-Schwarzschild black hole. A sphere on the regulated boundary encloses an infinitely large spatial region that extends all the way to the second, disconnected conformal boundary on the far side of the black hole.}
\label{fig-esch}
\end{figure}

Another example is furnished by the eternal Schwarzschild-AdS black hole solution.  Let us again consider the regulated global boundary given by the direct product of time with a sphere of fixed radius larger than the black hole, $b={\mathbf R}\times {\mathbf S}^{2}$. As shown in Fig.~\ref{fig-esch}, this sphere encloses an infinitely large spatial region that extends all the way to the second, disconnected conformal boundary on the far side of the black hole.  This region contains arbitrarily large entropy: for example, a dilute gas of $n\to\infty$ photons can be added near the second boundary with negligible backreaction.  Of course, our intuition tells us that the CFT that lives on the second boundary should be ``responsible'' for those photons.  The only problem is that spacelike holography does {\em not\/} tell us this.  In fact, it cannot possibly tell us any such thing. Black hole event horizons are global objects; for example, if we decide much later to add mass to the black hole, this will affect the location of the horizon at all earlier times.   Thus, there is no local criterion that can prevent us from extending a spacelike holographic domain to the far side of the black hole.

As the above examples illustrate, the light-like nature of the holographic relation between entropy and area is crucial for making sense of AdS/CFT as a manifestation of the holographic principle.  This relation is captured by the geometric construction of light-sheets.

\subsection{Bounds on the Holographic Domain}
\label{sec-light}

As shown by Susskind and Witten~\cite{SusWit98}, the CFT with a UV cutoff $\delta$ has a maximum entropy equal to the proper area of the spatial boundary of AdS in standard global coordinates.  The latter is considered to be located in the bulk, a coordinate distance $\delta$ away from the true conformal boundary.  This observation tells us that the CFT manifestly has the correct number of degrees of freedom demanded by the holographic principle, given by the area of the boundary, not by the size of the enclosed volume.  

In this argument one assumes, of course, that the CFT describes no more and no less than the spacetime region ``enclosed'' in the regulated boundary.  This is plausible in the case of global AdS with the standard slicing of the boundary into round spheres.  The analysis of the previous subsection has shown, however, that this assumption is coordinate-dependent at best, and that it is ill-defined for cases such as the Poincar\'e patch, where the boundary slices do not ``enclose'' any particular region.

In order to claim more generally that the CFT uses an area's worth of degrees of freedom to describe the bulk, we must characterize some relevant bulk region, given the boundary.  Note that the UV/IR connection is not the issue; by construction, the Susskind-Witten argument suffices to ensure that the maximum entropy on the CFT agrees with the area of the regulated boundary.  This remains true for arbitrary foliations of the boundary, as long as the short-distance regulator in the CFT is imposed with respect to the chosen time-slices.  We can always remove the UV cutoff at the end and think of the CFT as living on the true conformal boundary.  

The nontrivial question is how far from the boundary (how deep into the bulk) the CFT description is valid.  If this region is taken to extend too far from the boundary, then the bulk entropy might be larger than the boundary area, and thus larger than the maximum entropy of the CFT.  In that case, the CFT Hilbert space would be too small to capture the bulk physics.  The only way to ensure that the bulk entropy is sufficiently small is to appeal to the covariant entropy bound.  This is why the relevant bulk region must be related to the boundary by light-sheets.  We will now explore possible concrete proposals for this relation.

Consider a timelike hypersurface $b$ embedded in an asymptotically AdS and a foliation into spacelike hypersurfaces.  We may view $b$ as a spacetime in its own right, with one less spatial dimension than the AdS spacetime it is embedded in.  In order for the theory living on $b$ to be well-defined, we require that $b$ be globally hyperbolic and that each time slice be a Cauchy surface.  Most relevant for the present discussion is the limiting case where $b$ lies on the conformal boundary of global AdS: $b\subset {\cal I}$. ($b$ can be a proper subset of ${\cal I}$, for example if $b$ is the boundary of the Poincar\'e patch of AdS.)  In this case, the metric of $b$ is defined only up to conformal transformations.  We will be interested only in situations where each time slice of $b$ is ``normal'', i.e, each slice $t$ admits both a past-directed light-sheet $\ell^-_t$ and a future-directed light-sheet $\ell^+_t$.  This is automatically the case for $b\subset {\cal I}$, as we show in appendix~\ref{sec-normal}.

Let us consider the regions $L^+(b)$ and $L^-(b)$.  $L^+(b)$ is defined as the union over $t$ of the future-directed light-sheets $\ell^+_t$ of each slice. Similarly, $L^-(b)$ is the union of all past-directed light-sheets. Two natural looking possible bounds on the holographic domain are $H(b) \subset L^+(b)$ and $H(b) \subset L^-(b)$. At the fundamental level, there is no distinction between the past and the future, so $L^+(b)$ and $L^-(b)$ should play a symmetric role as bounds on the holographic domain.  Let us therefore consider the candidate bounds $H(b) \subset L^+(b) \cup L^-(b)$, and $H(b) \subset L^+(b) \cap L^-(b)$. The former bound guarantees that every point in $ L^+(b) \cup L^-(b)$ lies either on a future-directed or a past-directed light-sheet from some time slice of $b$.   The latter bound is stronger: if $H(b) \subset L^+(b) \cap L^-(b)$, then every point in $H(b)$ lies on at least one past- and one future-directed light-sheet.

The choice of time slicing on $b$ is a coordinate choice and so cannot have fundamental significance. Remarkably, the set $L(b) \equiv L^+(b)\cap L^-(b)$ is indeed independent of how $b$ is foliated, even though $L^+(b)$ and $L^-(b)$ individually do depend on the time slicing of $b$. In the following section we will prove a stronger theorem: $L(b)=C(b)$.  (Recall that $C(b)\equiv J^+(b)\cap J^-(b)$ is the region causally connected to $b$, which we discussed in the previous section.)  The fact that $L(b)$ is slicing-independent follows as a corollary, since $C(b)$ is slicing-independent by construction. The simplicity and slicing-independence of $L(b)$ make it especially attractive, and we will see in Sec.~\ref{sec-RG} that it also leads to a useful formulation of holographic RG flow.

\begin{figure}[tbp]
\centering
\subfigure{
\includegraphics[width=1.75in]{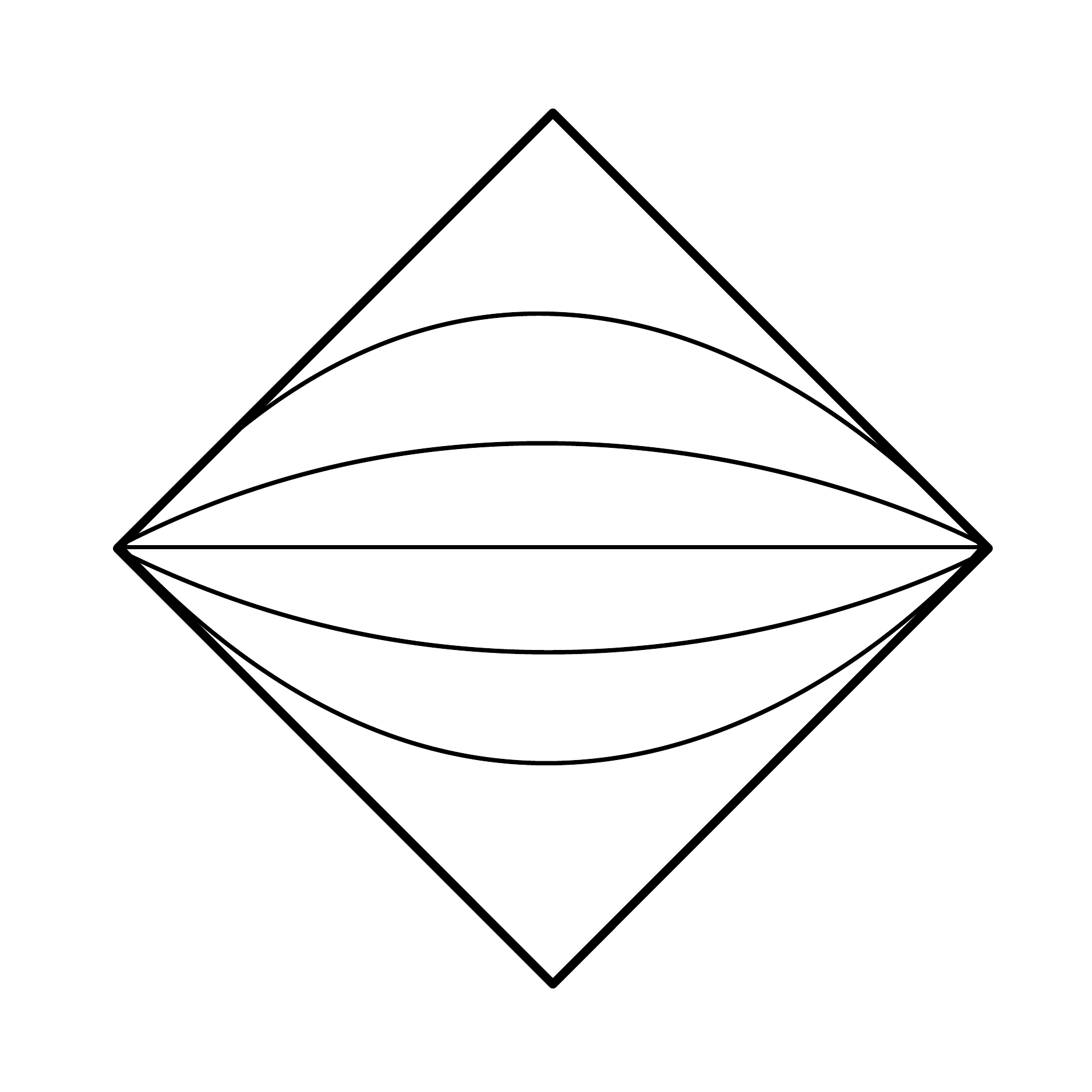}
}
\subfigure{
\includegraphics[width=2in]{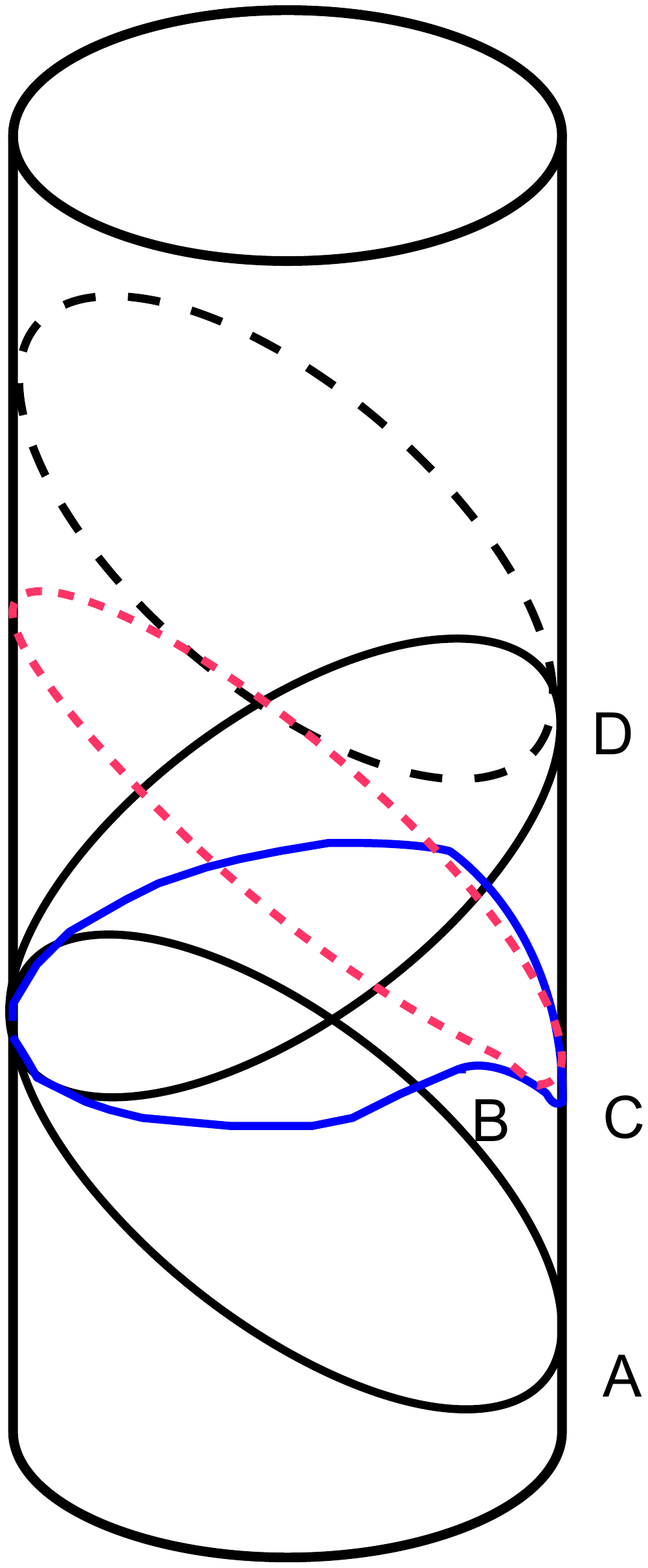}
}
\subfigure{
\includegraphics[width=1.75in]{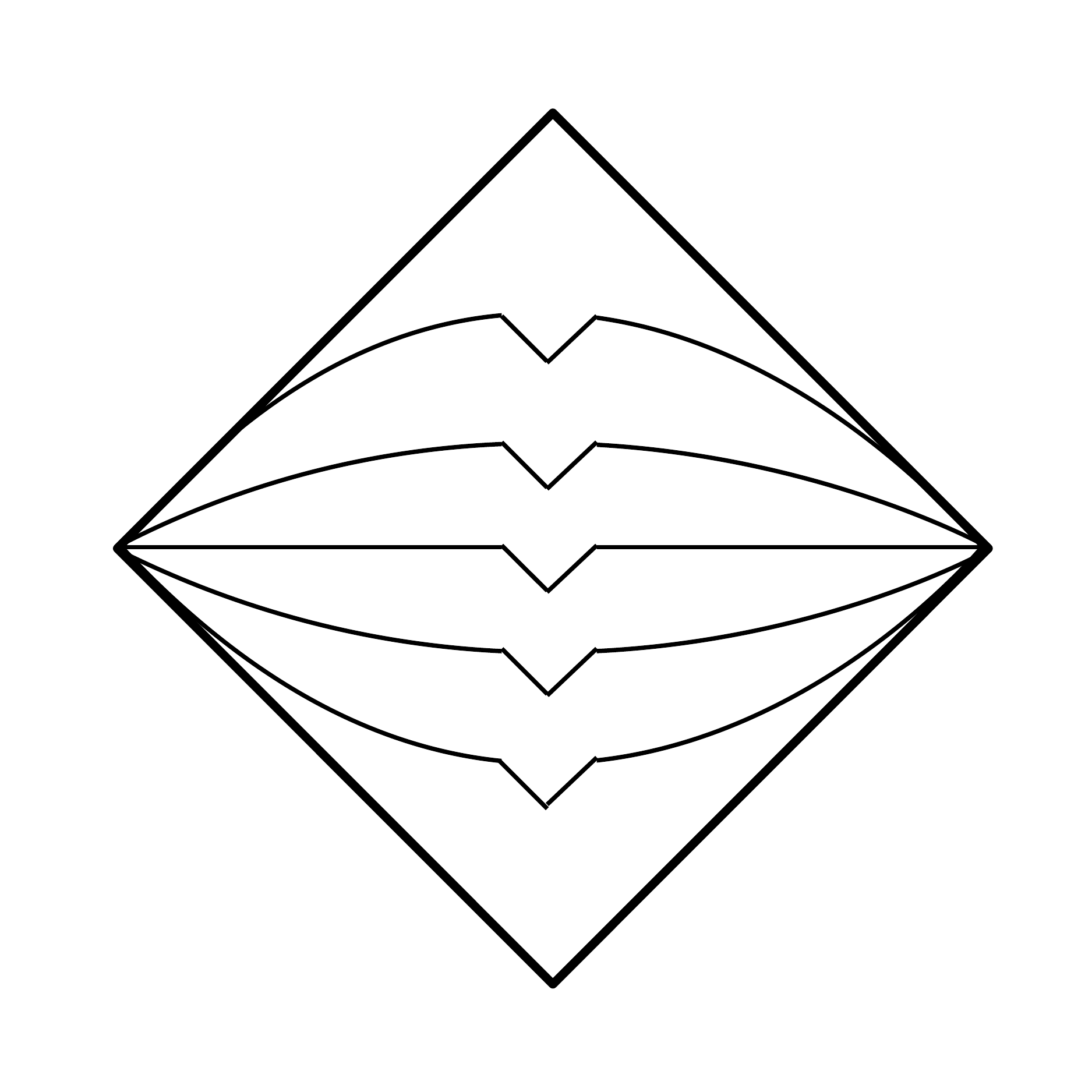}
}
\caption{The union of all future directed light-sheets, $L^+(b)$, coming off the usual slicing of the boundary Minkowski space (left) covers precisely the Poincar\'e patch (the wedge-shaped region that lies both in the future of the boundary point $A$ and in the past of $D$). On the right, we show a a different time slicing of the same boundary region. One of these slices is shown in blue in the bulk Penrose diagram (center); it curves up at $B$ and down at $C$. The future-directed light-sheet coming off the portion of the slice near $C$ is nearly the same as the future lightcone of $C$ (shown in red/short-dashed), which reaches far beyond the Poincar\'e patch to the far side of AdS.  The bulk region covered by $L^+(b)$ will thus be nearly two Poincar\'e patches, consisting of the points that lie in the future of $A$ but not in the future of $D$ (long-dashed).}
\label{fig-union}
\end{figure}

The other set that we were led to consider, the union $L^+(b) \cup L^-(b)$, is slicing-dependent (see Fig.~\ref{fig-union}). This disqualifies it from further consideration, if we insist, as we do in this paper, that a unique upper bound on the holographic domain should be constructed from light-sheets off a single, arbitrary slicing of the boundary region $b$.   

It is important to note that there are other, more complicated ways of constructing a bulk region covariantly from light-sheets. We will not analyze such sets here, but we mention some of them for completeness and future consideration.  For example, to treat the possible time-slicings in a democratic way, one could form the union, or the intersection, over all possible time-slicings $\mathcal{T}$, of the sets $L^+(b) \cup L^-(b)$: $L'(b) = \bigcap_{\mathcal T} (L^+(b) \cup L^-(b))$ and $L''(b) = \bigcup_{\mathcal T} (L^+(b) \cup L^-(b))$. Clearly, $L(b) \subset L'(b) \subset L''(b)$.  However, the restriction $H(b)\subset L''(b)$ seems too weak to ensure that the boundary theory has enough degrees of freedom to describe the bulk.  For a given time slicing, it may not be the case that a given point in $L''(b)$ lies on any light-sheet emanating from a slice.  Worse, there may not exist any choice of time slicing for which all of $L''(b)$ is covered by the light-sheets from the slices. The stronger bound $H(b) \subset L'(b)$ (which however is weaker than the bound we examine here) does ensure these properties, and we intend to investigate it further in future work.

In summary, the requirement that the bulk have no more degrees of freedom than the boundary, combined with arguments of symmetry and simplicity, has led us to propose the upper bound
\begin{equation}
L(b)\supset H(b)
\end{equation}
on the holographic domain, where $L(b)\equiv L^+(b)\cup L^-(b)$ and $L^+(b)$ ($L^-(b)$) is the union of all future-directed (past-directed) light-sheets of the time-slices that foliate the boundary region $b$.

\section{Proof that $L=C$}
\label{sec-equiv}

In the previous two sections we have argued that the holographic bulk domain $H$ dual to the boundary region $b$ must satisfy
\begin{equation}
C(b)\subset H(b)\subset L(b)~.
\label{eq-chl}
\end{equation}
In this section, we will prove that
\begin{equation}
C(b)=L(b)~.
\end{equation}
This implies that
\begin{equation}
C(b)= H(b)= L(b)~,
\label{eq-chl}
\end{equation}
so the holographic dual $H$ is completely determined by our assumptions.\footnote{For certain choices of $b$, $H$ thus excludes bulk points that can be represented on $b$~\cite{PolSus99,HeeMar}; see, however, Sec.~\ref{sec-nonlocal}.  Our result is consistent with the fact that only $C$ is needed to compute CFT correlation functions in $b$~\cite{Mar04}.}

It is obvious that $L(b)\subset C(b)$.  But the converse inclusion is nontrivial, particularly since $L(b)$ is constructed from two sets $L^\pm(b)$ that depend on the slicing of $b$, whereas $C(b)$ is slicing-independent.  By Eq.~(\ref{eq-chl}), $C(b)\subset L(b)$ is required for the consistency of the arguments we have put forward in the previous section.  Thus, our proof also serves as a nontrivial consistency check.  It is the main technical result of this paper.

We begin by stating our assumptions and definitions.  Let $B$ be a manifold with boundary, and let $b\subset \partial B$ be a timelike embedded submanifold in $B$.  We will require that $b$ is globally hyperbolic when considered as a spacetime on its own. We also assume that $B$ has the property that $J^\pm(P)$ is closed for every $P\in B$. Note that we are not assuming inextendibility of either $b$ or $B$, nor global hyperbolicity of $B$ (from which the latter assumption would follow).

In the application of our theorem to AdS/CFT, we take $b$ to be a portion of the conformal boundary of AdS.  In this case, the spacetime $B$ should be taken as the union of the unphysical conformally rescaled AdS spacetime and its boundary. Since our theorem only relies on properties of the spacetime which are preserved by conformal transformations, we are free to construct the proof in this unphysical spacetime.

An additional assumption is that the causal relation between any two points in $b$ computed according to the causal structure of the lower-dimensional spacetime is the same as that according to the causal structure of $B$ itself. This is essential for a physical duality to hold: the only way to guarantee that causality is preserved on both sides of the duality is to make the causal structures compatible in this way.  (In asymptotically AdS spacetimes, this assumption follows from a theorem of Gao and Wald~\cite{GaoWal00}.  However, this theorem relies on additional assumptions that we have no reason to make here.)

Let $\tau:b\to \mathbf{R}$ be an arbitrary time function\footnote{That is, $\tau$ is differentiable on $b$, and $\nabla^a \tau$ is a past-directed timelike vector field.} on $b$ such that the equal time slices are Cauchy surfaces of $b$. The existence of $\tau$ is guaranteed by theorem 8.3.14 of Ref.~\cite{Wald}. Let $K\subset \mathbf{R}$ be the image of $\tau$.  Note that $K$ is an open interval. Let $s_t \subset b$ be the Cauchy surface consisting of points with time $t$, $s_t = \left\{p\in b | \tau(p) = t\right\}$.  Note that each $s_t$ is spacelike. We will not demand that $\tau$ extends to a time function on $B$.

For the purpose of this theorem, we shall define $\ell^+_t \subset B$ ($\ell^-_t \subset B$) as the set of future-directed (past-directed) null geodesics which are orthogonal to $s_t$ with no conjugate points between $s_t$ and the endpoint. ($\ell^\pm_t$ is a light-sheet associated to $s_t$ if it is initially nonexpanding away from $s_t$.  For the case of interest, where $b$ is a portion of the conformal boundary of AdS, this always holds in the {\em physical\/} spacetime; see Appendix~\ref{sec-normal}.   In the proof we will not make use of the nonexpansion property and so will not demand that $\ell^\pm_t$ be a true light-sheet in the unphysical spacetime.) Let $L^\pm = \bigcup_t \ell^\pm_t$, and $L = L^+ \cap L^-$. 
\begin{figure}[tbp]
\centering
\includegraphics[width=3in]{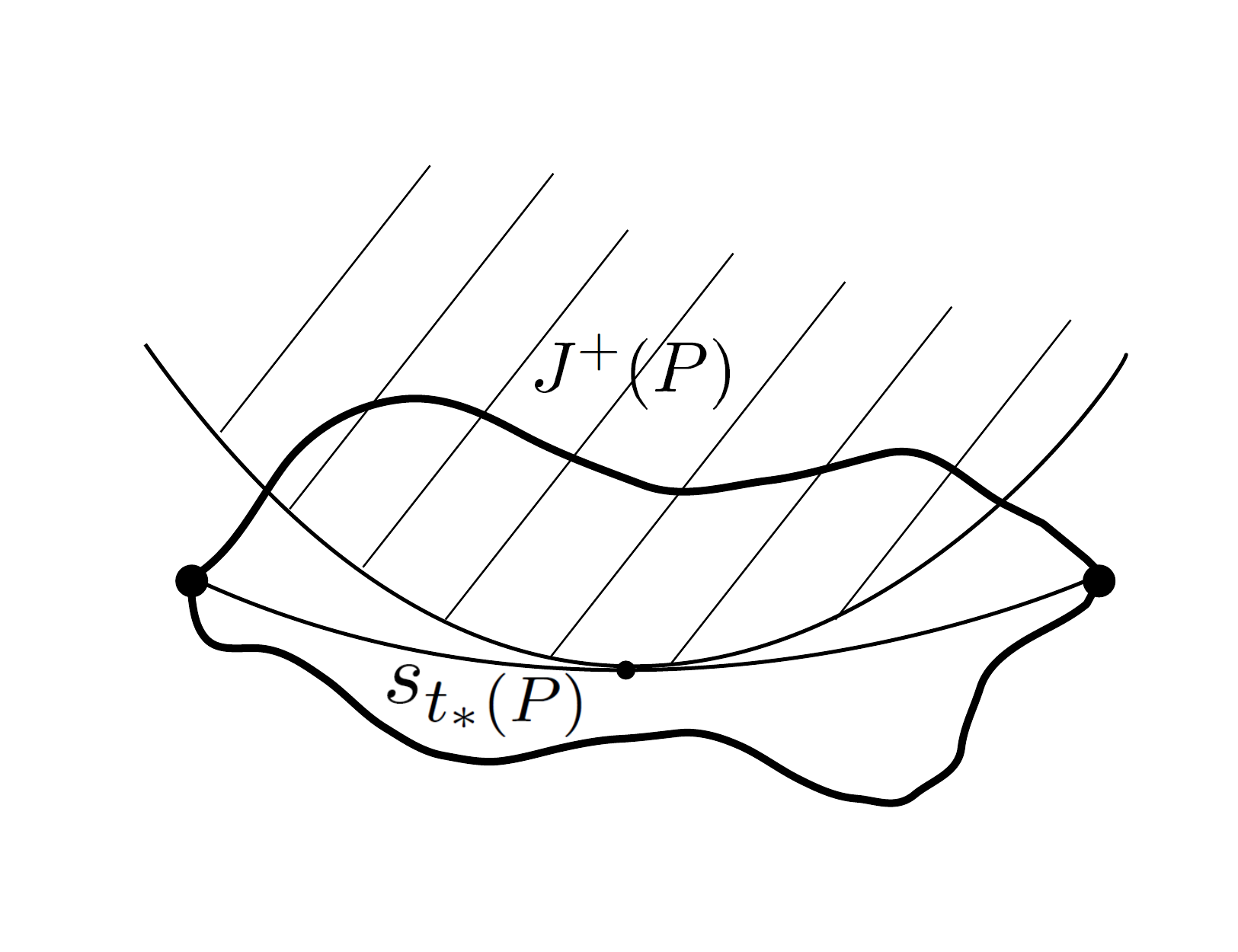}
\caption{Consider an arbitrary boundary region $b$ (enclosed by the solid line), and a point $P$ in the bulk region $C(b)$ (orthogonal to the page). The causal future of $P$, $J^+(P)$, intersected with $b$ is shown hatched. Roughly, the strategy of the proof is to demonstrate that there exists a time slice $s_{t_*(P)}$ on the boundary that is tangent to the lower boundary of future of $P$ in $b$.  We show that $s_{t_*(P)}$ is the earliest time slice that has any intersection with the future of $P$, and that $P$ lies on the past light-sheet of $s_{t_*(P)}$.}
\label{fig-proof}
\end{figure}

\paragraph{Theorem}  {\em $L=C$ for any choice of time function on $b$.}
where $C\subset B$ is the set of points $P$ that lie on a causal curve that begins and ends on $b$.  

\paragraph{Corollary} $L$ does not depend on the choice of time function, even though $L^+$ and $L^-$ do.

\paragraph{Proof} Trivially, $L\subset C$.  It remains to be shown that $C\subset L$, for all time functions $\tau$ (i.e., foliations of $b$).  We will show that $C\subset L^-$, and similar arguments show that $C\subset L^+$. Let $P\in C$ (and $P$ is not contained in $b$) .

{\em Definition.}  Let $K^+(P)$ be the subset of the real numbers consisting of all $t$ such that $s_t \cap I^+(P)$ is not empty.  That is, $\left\{s_t | t\in K^+(P)\right\}$ is the set of all time slices on $b$ which contain endpoints of future-directed timelike curves from $P$.

{\em Lemma A.} $K^+(P)$ is nonempty. {\em Proof.} Since $P\in C$, there is a point $p\in b \cap J^+(P)$. All points in $b\cap I^+(p)$ are necessarily in $I^+(P)$ (corollary following theorem 8.1.2 of Ref.~\cite{Wald}). Since $K$ is an open interval, there exists $t\in K$ with $t> \tau(p)$. Any inextendible timelike curve passing through $p$ intersects $s_{t}$ because $s_{t}$ is a Cauchy surface, which shows that $s_{t} \cap I^+(p)$, and consequently $b\cap I^+(P)$, is nonempty. Therefore $K^+(P)$ is nonempty.

{\em Lemma B.} For any $t \in K^+(P)$, the set $A_t = b\cap J^+(P) \cap J^-(s_t)$ is compact. {\em Proof.} Since $P\in C$, $b\cap J^-(P)$ is not empty. Let $q\in b \cap J^-(P)$. Let $j^+(q)$ be the causal future of $q$ within the spacetime $b$, and define $j^-(s_t)$ similarly. Then $j^+(q) \cap j^-(s_t)$ is compact (theorem 8.3.12 of Ref.~\cite{Wald}). But because the causal relation between points in $b$ is the same whether we treat them as events in the spacetime $b$ or the spacetime $B$, we also have that $j^+(q) \cap j^-(s_t) = b\cap J^+(q) \cap J^-(s_t)$. Since this set is compact, it is closed as a subset of $B$. $J^+(P)$ is a closed subset of $J^+(q)$, and it follows that $b\cap J^+(P) \cap J^-(s_t) = J^+(P) \cap(b\cap J^+(q) \cap J^-(s_t))$ is a closed subset of $b\cap J^+(q) \cap J^-(s_t)$. Therefore $b\cap J^+(P)\cap J^-(s_t)$ is compact.

{\em Definition.} Let $t_*(P) = {\rm inf}_{t\in K^+(P)}~t$.

{\em Lemma C.} For any $t\in K^+(P)$, $t_*(P) = {\rm min}_{p\in A_t}~\tau(p)$. In particular, this means that there is a surface $s_{t_*(P)}$ and that $s_{t_*(P)}\cap J^+(P)$ is not empty. {\em Proof.} By Lemma B, $A_t$ is compact and hence $\tau(p)$ attains a minimum value $\tau_{\rm min}$. Consider a point $p\in s_{\tau_{\rm min}} \cap J^+(P)$ and an inextendible future-directed timelike curve $\gamma$ in $b$ starting at $p$. All points on $\gamma$ other than $p$ are necessarily in $b\cap I^+(P)$ (corollary following theorem 8.1.2 of Ref.~\cite{Wald}), and $\gamma$ necessarily intersects $s_{t'}$ for every $t'\in K$ with $t'>\tau_{\rm min}$ because $s_{t'}$ is a Cauchy surface. This implies that all such times $t'$ are in $K^+(P)$, and hence $\tau_{\rm min} \geq t_*(P)$. If $\tau_{\rm min} > t_*(P)$, then there is some $t' \in K^+(P)$ with $\tau_{\rm min} > t'$. Then $t>t'$, and so $s_{t'} \cap A_t$ is not empty by the definition of $A_t$. But then $\tau_{\rm min} \leq t'$ by the definition of $\tau_{\rm min}$, which contradicts $\tau_{\rm min} > t'$. Therefore we conclude that $\tau_{\rm min} = t_*(P)$.

{\em Lemma D.} $K^+(P)$ is an open subset of the real numbers.  This implies that $s_{t_*(P)} \cap I^+(P)$ is empty. {\em Proof.} Let $\gamma_p$ be an inextendible timelike curve in $b$ which passes through a point $p\in b\cap I^+(P)$. We can choose to parametrize $\gamma_p$ by the time function $\tau$, which means that $\gamma_p:K\to b$ is a continuous function satisfying $\tau(\gamma_p(t)) = t$. Since $I^+(P)$ is open, it follows that the inverse image $\gamma_p^{-1}[I^+(P)]$ is open in $K$, and therefore open in $\mathbf{R}$ because $K$ is an open interval. By applying $\tau\circ\gamma$, we see that $\gamma_p^{-1}[I^+(P)] \subset K^+(P)$. Then  $\bigcup_{p\in b\cap I^+(P)}  \gamma_p^{-1}[I^+(P)]$ is an open subset of $\mathbf{R}$ equal to $K^+(P)$.

{\em Proof of theorem.} Lemma C and Lemma D together demonstrate that there is a surface $s_{t_*(P)}$ such that $s_{t_*(P)}\cap I^+(P)$ is empty while $s_{t_*(P)} \cap J^+(P)$ is nonempty. Let $p\in s_{t_*}(P)\cap J^+(P)$. By the corollary following theorem 8.1.2 of Ref.~\cite{Wald}, there is a null geodesic connecting $p$ to $P$. Furthermore, since $s_{t_*(P)} \cap I^+(P)$ is empty, this null geodesic cannot be deformed to a timelike curve connecting $s_{t_*(P)}$ to $P$. Then by theorem 9.3.10 of Ref.~\cite{Wald}, this null geodesic must be orthogonal to $s_{t_*(P)}$ and have no conjugate points between $P$ and $p$. This shows that $P\in \ell^-_{t_*(P)} \subset L^-$.

\section{Covariant Renormalization Group Flow}
\label{sec-RG}

We repeatedly made use of the UV/IR connection~\cite{SusWit98} in motivating our constructions. Entropy bounds play a very important role in UV/IR: the bulk region within the IR cutoff must have entropy limited by the area in Planck units of a time slice on the cutoff surface. However, as we have stressed, this only holds true for very special time slices and the covariant entropy bound must be used in general to bound the bulk entropy. Holographic renomalization group flows~\cite{BalKra99, DebVer99, FAuLiu10, HeePol10} aim to refine UV/IR, but in all standard approaches manifest covariance is lost and the status of entropy bounds is unclear. Here we outline an approach which reproduces the standard results and remedies both of these problems. Our construction gives an improved bulk radial flow, however it does not address the open question of finding a precise field theory RG representation of the bulk flow. 
\begin{figure}[h]
\centering
\subfigure{
\includegraphics[width=2in]{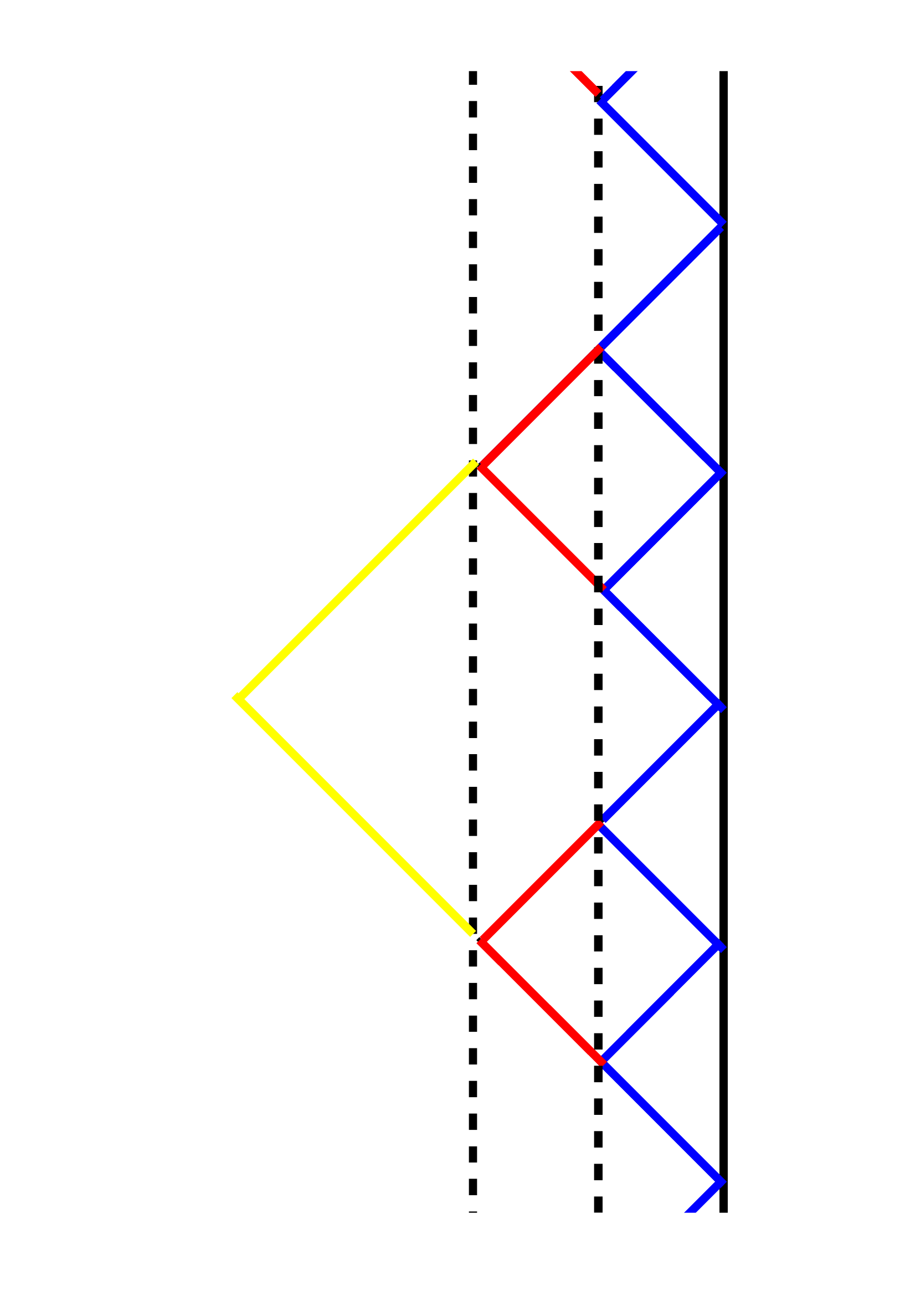}
}
\subfigure{
\includegraphics[width=2in]{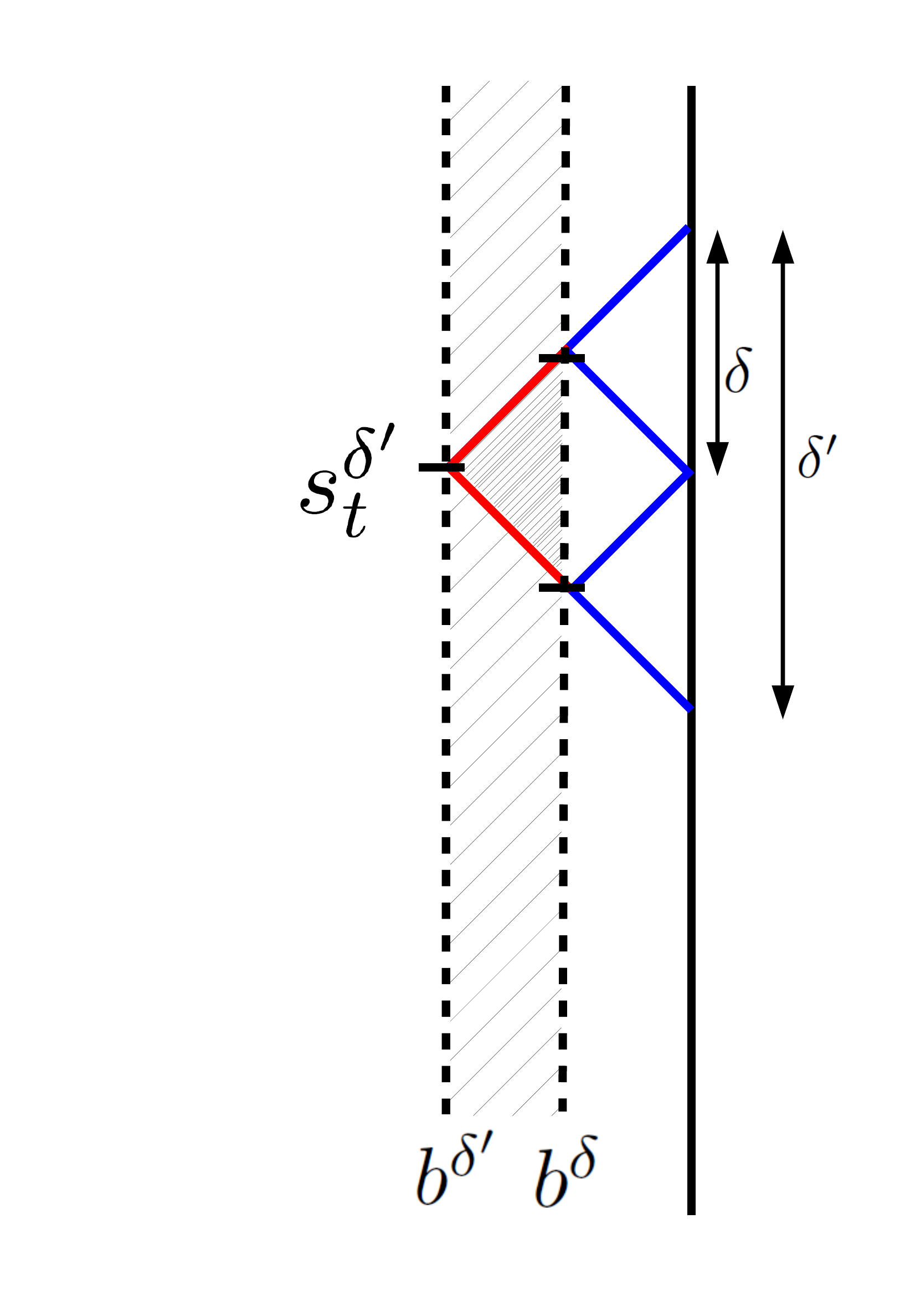}
}
\caption{The covariant bulk RG flow presented here reproduces the standard bulk RG flow in certain coordinate systems. Here we illustrate the construction in global coordinates of Anti-de Sitter space.  For a given coordinate time cutoff $\delta$, the union over $t$ of the intersection surfaces $s_t^\delta = \ell^-_{t+\delta/2} \cap \ell^+_{t-\delta/2}$ form a timelike hypersurface $b^\delta$ in the bulk (left).  The cross-sectional area of a given light-sheet will be greater on the surface $b^{\delta}$ than on $b^{\delta'}$ (right). The difference $(A-A')/4$ bounds the entropy on the red light-sheets going from $b^{\delta}$ to $b^{\delta'}$, meaning that the bound applies to the entire darkly shaded wedge between them. The lightly shaded region between the hypersurfaces $b^{\delta'}$ and $b^{\delta}$ is covered by such wedges.}
\label{fig-RG}
\end{figure}

Choose a time function $\tau$ on $b$. Then for any $t_1$, $t_2$ the set $b[t_1,t_2] = \{p\in b | t_1 < \tau(p) < t_2\}$ satisfies the conditions of our theorem and we can associate to it the region $H(b[t_1,t_2])$.  Now introduce a cutoff timescale $\delta$ to the theory.  In the bulk, we should remove the union (over $t$) of the sets $H(b[t-\delta/2,t+\delta/2])$ from $H(b)$. The remaining bulk region, $H^\delta$, is the region described by the cutoff CFT. This prescription is similar to the construction of~\cite{Bou09}, where it was shown that the IR cutoff surface as normally defined can be reproduced using only causality.

Denote the boundary of $H^\delta$ by $b^\delta$. The time function on $b$ automatically induces a time function on $b^\delta$:\footnote{This will be true for sufficiently well-behaved time functions $t$; the precise conditions will be investigated elsewhere.} By construction, $b^\delta$ is the union of sets of the form $s^\delta_t = \ell^-_{t+\delta/2} \cap \ell^+_{t-\delta/2}$, which we can take as time slices on $b^\delta$.  The maximum proper time in $b^\delta$ between $s^\delta_t$ and $s^\delta_{t+\delta}$ will be of order the AdS time. More generally, we can let $\delta$ depend on $t$.  

To change the cutoff from $\delta$ to $\delta'$, we have two equivalent options: First, we can return to the true boundary $b$ and repeat the construction with $\delta'$ in place of $\delta$. Second, we can use the surface $b^\delta$, together with its induced time function, as the starting point for the construction, with cutoff $\delta'-\delta$. The geometry is the same either way, because the light-sheets from $b^\delta$ are continuations of the light-sheets from $b$.

By the definition of the $L$, and since $H=L$, all time slices of $b^\delta$ are {\em normal}.  That is, their orthogonal light-rays are everywhere nonexpanding in the direction of the bulk RG flow.  A generic time slicing on $b^\delta$ will {\em not\/} admit ingoing past- and future-directed light-sheets at all points on all time-slices; only the slicing induced by the flow has this property.  And generic hypersurfaces other than those induced by the flow may not admit any slicing with this property. Note that this property is inherited from the remarkable property of the conformal boundary described in the Appendix: any slicing of the conformal boundary is everywhere normal in the physical metric.

The fact that all time slices of $b^\delta$ are normal is highly nontrivial.  It leads to two additional, attractive features that distinguish this geometric flow from, say, the flow along spacelike geodesics:
\begin{itemize}
\item Entropy bounds hold for both the UV and the IR regions. 
\item These bounds guarantee that the effective number of degrees of freedom is nonincreasing along the flow.
\end{itemize}

Consider first the IR region.  The covariant entropy bound guarantees that any future- or past-directed light-sheet from any slice $s^\delta_t$ has entropy less than the area of that slice, in Planck units.  Since the area of the light-sheet is nonincreasing as it moves away from the boundary,  the number of degrees of freedom is nonincreasing along the RG flow.

Now consider the UV region, i.e., follow the light-sheet from $s^{\delta'}_t$ backward to some cutoff $\delta<\delta'$ closer to the true boundary.  The cross-sectional area will be larger on $b^{\delta}$ and we can consider the area difference $A-A'$.  A generalized version of the covariant entropy bound~\cite{FMW} implies that the entropy on the {\em partial light-sheet} between $A$ and $A'$ is bounded by $(A-A')/4$.  Note that this bound applies to both light-sheets bounding the wedge-shaped region between the slices  $s^{\delta}_{t+(\delta'-\delta)/2}$ and $s^{\delta}_{t-(\delta'-\delta)/2}$ on $b^{\delta}$ and $s^{\delta'}_t$ on $b^{\delta'}$, and hence to the entire information content of the regions that are integrated over in the RG flow.\footnote{In general, the area $A(t)$ of different slices on $b^{\delta}$ will not be independent of $t$.  It seems likely that requiring this independence will lead to preferred choices of $\delta(t)$ and thus for the flow.  Also, while the areas will be automatically finite at finite $\delta$ in the case of the global boundary, they will be infinite in general and must be regulated.}

RG flows are normally defined in terms of proper distances, times, or energies, whereas the above construction is in terms of an arbitrary coordinate time. To define proper distances on the boundary, one has to choose a conformal frame, and the choice of time function in our construction is analogous. Heuristically, one can imagine choosing the conformal factor so that $\tau$ becomes proper time, which means that it is conjugate to energy. 

We have emphasized the geometric aspects of the RG flow, i.e., the bulk side.  The question of what precisely the removal of near-boundary regions corresponds to in the CFT remains subtle, and it will be interesting to revisit it in light of the geometric flow we have described.  In particular, it would be nice to understand whether a duality holds for the wedge-shaped regions associated with thin boundary strips, and for the Rindler portion of the bulk dual to small causal diamonds on the boundary.  We are currently studying this question~\cite{BouFreTA}; see also~\cite{HamKab05, HamKab06, Kab11}.  

As the example of the global boundary (Fig.~\ref{fig-RG}) illustrates, the description of the bulk region $H(b[t-\delta/2,t+\delta/2])$, for small time intervals $\delta$, cannot involve the full set of CFT degrees of freedom, since approximately $\exp(A/4)$ CFT states correspond to bulk states localized to the interior of the bulk hypersurface $b^\delta$ .  If a duality exists, the CFT degrees of freedom relevant for $H(b^\delta)$ will not involve certain nonlocal operators that occupy scales larger than $\delta$.

\section{Nonlocal Operators and the Bulk Dual}
\label{sec-nonlocal}

We constructed a candidate for a bulk region $H(b)$ dual to $b$ geometrically, using considerations of causality and holographic entropy bounds.  Except for appealing to the UV/IR relation~\cite{SusWit98}, we did not make use of detailed properties of the AdS/CFT duality or the boundary theory.  In this section we will explore a different approach to this question, namely the use of nonlocal operators in the CFT to probe deeply into the bulk region.  We will examine the relationships between this approach and the geometric one.  Note that the operator approach is available only for choices of $b$ where the construction of boundary duals to local bulk operators is known, so it is less general than our geometric construction.  

The methods of the operator approach were recently discussed by Heemskerk, Marolf, and Polchinski~\cite{HeeMar}, and we refer the reader to that work for further details. The problem is to identify a subset of operators $\{\phi(x)\}$ within the set of all operators in the CFT on $b$, indexed by a position $x$ in a semiclassical spacetime of higher dimension, which can be identified as local bulk operators.  An important point is that the definition of the local bulk operators depends on the background metric.  Our analysis below pertains to the case where this metric is held fixed up to perturbative corrections in $1/N$, so this issue should not pose difficulties.

\begin{figure}[h]
\centering
\includegraphics[width=3in]{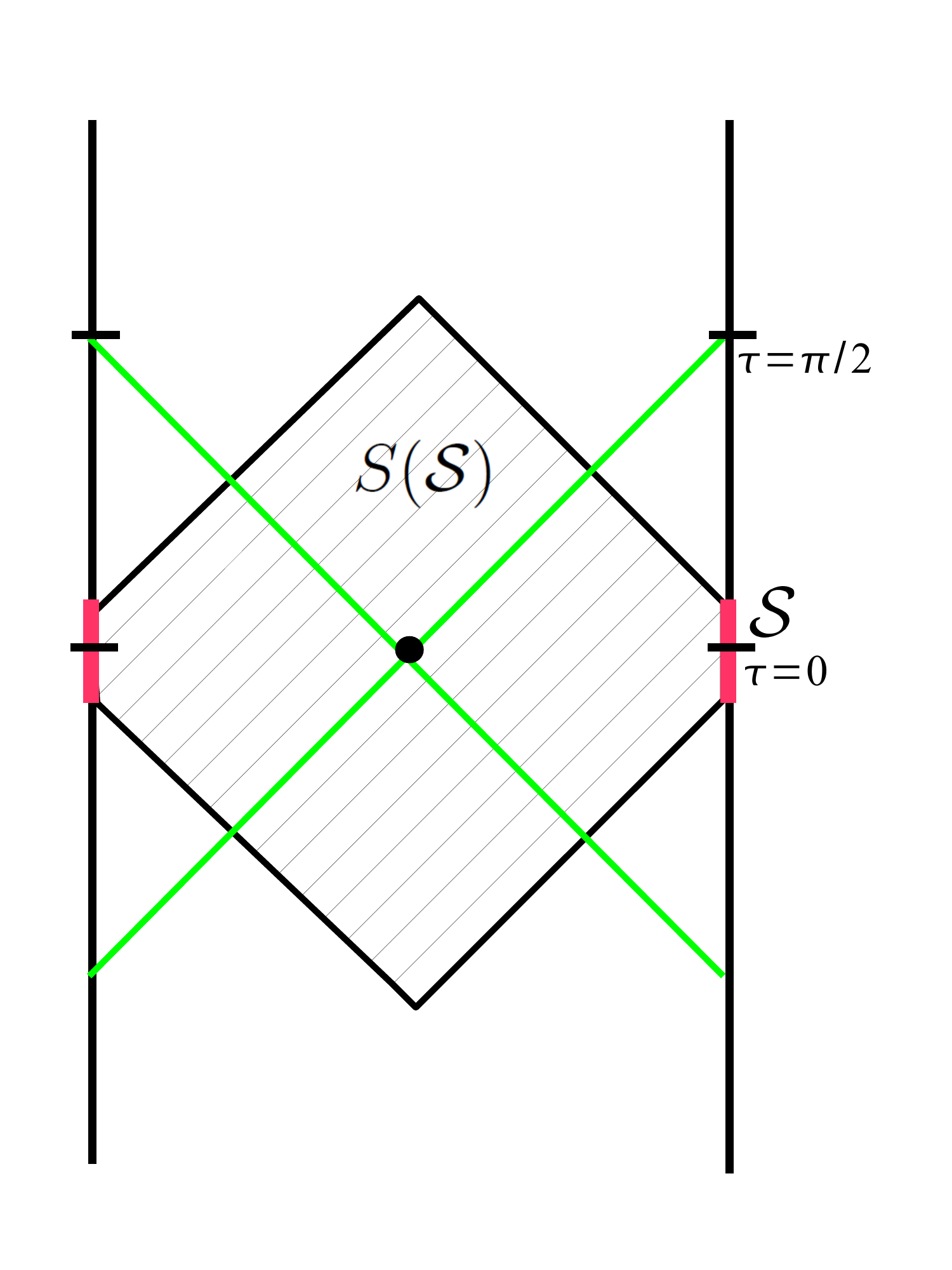}
\caption{A cross-section of Anti-de Sitter space, showing a short strip region $\mathcal{S}$ centered around $\tau=0$ on the boundary, and the bulk region $S(\mathcal{S})$ spacelike separated from $\mathcal{S}$. A local operator at the origin of the bulk can be written in terms of local operators on the boundary smeared over the boundary region spacelike-related to the origin, within the green wedges.  This region is much larger than $\mathcal{S}$ (red thick line), stretching from $\tau=-\pi/2$ to $\tau=+\pi/2$.}
\label{fig-smearing}
\end{figure}

The CFT definition of $\phi(x)$ will involve nonlocal CFT operators known as precursors~\cite{PolSus99,Sus99,Fre02}. An (in principle) explicit construction is available in the case of global AdS, and the resulting operators make use of an entire Cauchy surface of the global boundary. For this reason, and also for simplicity, we will spend most of this section focused on the case of a short strip $b=\mathcal{S}$,
\begin{equation}
\mathcal{S} = (-\tau_0,\tau_0)  \times S^d,  ~\text{with} ~\tau_0 \ll 1~,
\end{equation}
which we normally think of as being embedded in the global boundary (see Fig.~\ref{fig-smearing}), using the coordinates of Eq.~\ref{eq:AdSGlobal}.  From the global point of view, the set of operators on the Cauchy surface $\tau=0$ of the boundary is complete, and so in particular contains the operator $\phi(x)$ for every point $x$ in the entire global bulk. 

Now consider modifications of the CFT Hamiltonian $\mathcal{H}$ outside $b$, and let us define $\bar H$ as the largest bulk region such that operators $\phi(x)$ for $x \in \bar H$ can be represented in terms of CFT operators in $b$ in an unambiguous way. There are some modifications of the CFT Hamiltonian which continue to yield a well-defined bulk Hamiltonian. For instance, we can insert a local source in the CFT whose effect in the bulk is to cause a particle to propagate causally inward from the boundary. As we discuss below, certain modifications of the CFT Hamiltonian which make use of nonlocal operators can lead to ambiguities in the bulk Hamiltonian. In the case where we allow only those modifications leading to well-defined bulk Hamiltonians, we identify a region $S$ on which the operators $\phi(x)$ have unambiguous CFT representations in $b$. In the more general case of modifications which lead to an ambiguous bulk Hamiltonian, we will identify $\bar H$ as the subset of $S$ for which $\phi(x)$ still has an unambiguous CFT representation in $b$.  We will note that $\bar H$ is closely related to $H$.

\paragraph{The region $S$} The only known construction of $\phi(x)$ in the boundary theory consists of two steps and applies either in global AdS or the Poincar\'e patch. In the first step, one writes a local bulk operator $\phi(x)$ as a smeared local operator on the boundary:
\begin{equation} \label{eq:smear}
\phi(x) = \int{dy' K(x|y')\mathcal{O}(y')} + O(1/N)~.
\end{equation}
Here $y'$ denotes a boundary coordinate, while $x$ is the bulk coordinate.\footnote{The presence of a source will modify this equation in the appropriate way given by Green's identity.} The smearing function $K(x|y)$ is not unique, but a convenient choice is nonzero if and only if $y$ spacelike-related to $x$~\cite{HamKab05, HamKab06, Kab11}. One can think of $K(x|y)$ as providing the solution to the equation of motion for $\phi(x)$ given boundary data on the asymptotic boundary; this is a spacelike analogue of the standard initial value problem.  Note that for any choice of $x$ in global AdS, the boundary support of $K(x|y)$ is larger than the region $\mathcal{S}$. The second step of the construction uses unitary evolution in the CFT to write all of the local CFT operators $O(y')$ appearing in Eq.~\ref{eq:smear} in terms of nonlocal operators defined at $\tau=0$ (or some other timeslice within the short strip $\mathcal{S}$).

Now let us identify a subset $S(\mathcal{S})$ of the bulk such that the above construction of the CFT operator $\phi(x)$ is independent of modifications of the boundary Hamiltonian outside $b$,\footnote{We impose the boundary condition that the boundary state at $\tau=0$ remain unchanged.} provided that these modifications continue to lead to a well-defined bulk Hamiltonian.   In Ref.~\cite{HeeMar}, it was shown that $S$ consists of the bulk points that are neither in the causal future of the future boundary of $\mathcal{S}$, nor in the causal past of its past boundary.  We will refer to such points somewhat loosely as being ``spacelike-related'' to $\mathcal{S}$ (see Fig.~\ref{fig-smearing}).  

At first, this result may seem surprising, so it is worth reviewing the argument for it~\cite{HeeMar}.  First, fix a fiducial CFT Hamiltonian $\mathcal{H}$ over the entire global boundary, and follow the above procedure to construct an operator $\mathcal{O}$ satisfying $\phi(x)=\mathcal{O}$, for some $x$ spacelike-related to $\mathcal{S}$ (the special case $x=0$ is pictured in Fig.~\ref{fig-smearing}). But there is another way we can produce the operator $\mathcal{O}$: We can use causal bulk evolution to write $\phi(x)$ in terms of $\phi(x')$ for points $x'$ in the future of the $x$:
\begin{equation}\label{eq:causal}
\phi(x) =  \int_\Sigma dx' \left[\phi(x')\nabla G_{\rm adv}(x|x') - \nabla \phi(x')G_{\rm adv}(x|x') \right]+ O(1/N)~,
\end{equation}
where $\Sigma$ consists of a bulk Cauchy surface and, possibly, a portion of the global boundary. We can use Eq.~\ref{eq:smear} to write the $\phi(x')$ appearing here in terms of operators on the boundary. By evolving $\phi(x)$ sufficiently far forward into the future, the $y'$-support of $K(x'|y')$ will lie in the region $\tau >0$ for every $x'$ appearing in Eq.~\ref{eq:causal}. Now we evolve this new smeared operator back to $\tau=0$ to obtain a second operator $\mathcal{O}'$. However, $\mathcal{O}' = \phi(x) = \mathcal{O}$, and so these two procedures actually give the same answer.  Now suppose we modify the CFT Hamiltonian from our fiducial choice $\mathcal{H}$ to $\tilde{\mathcal{H}}$, and we stipulate that $\tilde{\mathcal{H}}$ only differs from $\mathcal{H}$ for $\tau<0$. We can repeat the procedure to compute new operators $\tilde{\mathcal{O}}$ and $\tilde{\mathcal{O}}'$, which are equal to each other and to $\phi(x)$. The claim is that, since the calculations of $\mathcal{O}'$ and $\tilde{\mathcal{O}}'$ refer only to the $\tau>0$ region of the bulk and boundary, and $\mathcal{H}=\tilde{\mathcal{H}}$ in that region, the computations are identical and so manifestly we have $\mathcal{O}' = \tilde{\mathcal{O}}'$. Therefore $\tilde{\mathcal{O}} = \mathcal{O}$, and the change in fiducial Hamiltonian did not change the operator assignment. An analogous argument can be made for modifications to the CFT Hamiltonian for $\tau>0$.
\begin{figure}[h]
\centering
\includegraphics[width=3in]{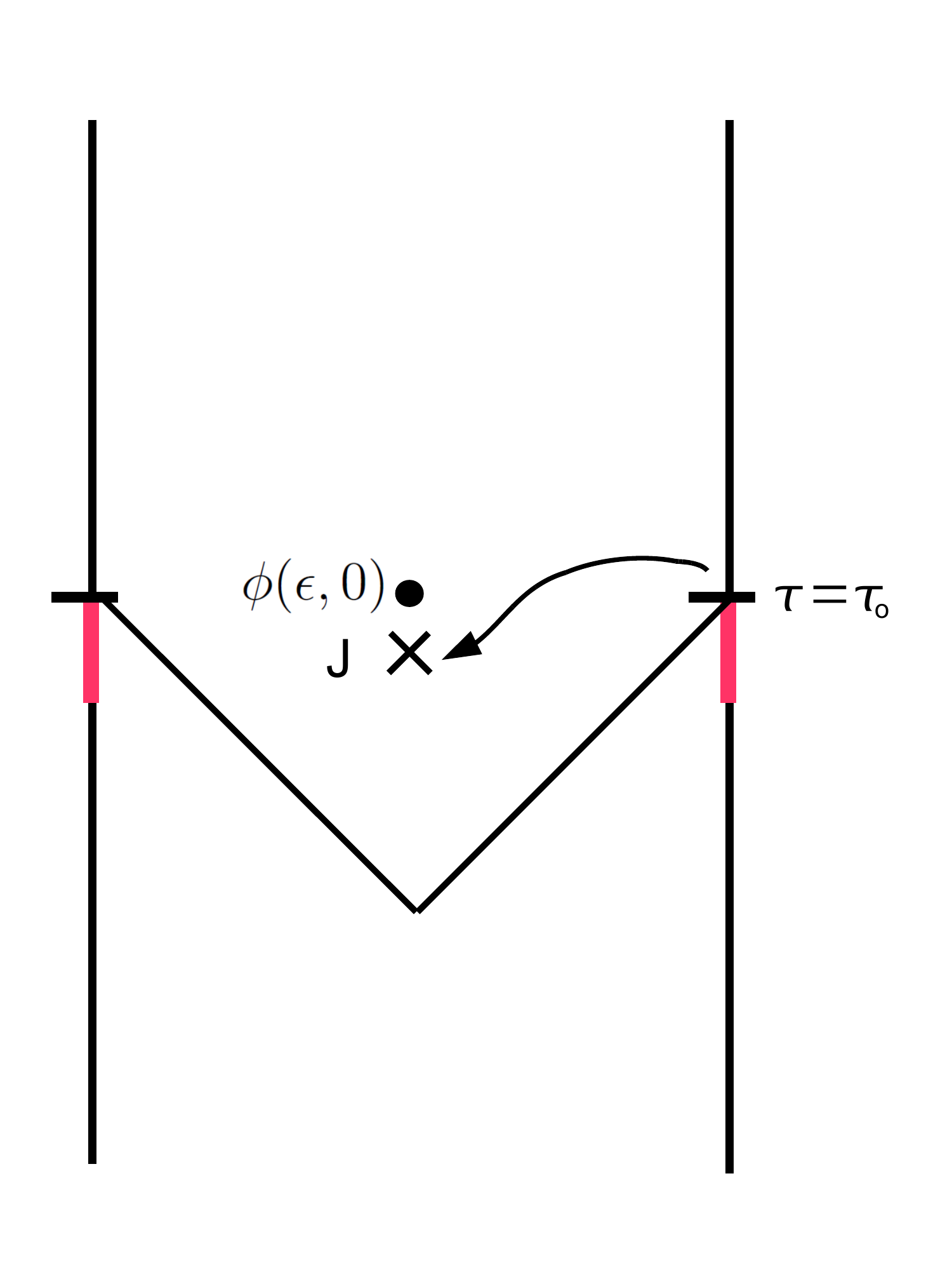}
\caption{According to the Hamiltonian on the boundary strip $ \mathcal{S}$, no source acts at the origin in the bulk, so the expectation value of $\phi$ vanishes everwhere. At the time $\tau_0$ outside the strip, a source term for the nonlocal boundary operator dual to $\phi(0,0)$ can be added to the boundary Hamiltonian. This causes the expectation value of $\phi$ to be nonzero in the future of $(0,0)$, in contradiction with the earlier conclusion about the same bulk points.  Thus, unless we possess information about the exterior of $\mathcal{S}$ on the boundary guaranteeing that such operators do not act, the bulk interpretation of regions outside $\bar H(\mathcal{S})=C(\mathcal{S})$ is potentially ambiguous.}
\label{fig-ambiguous}
\end{figure}

\paragraph{The region ${\bar H}$}

The above argument relied crucially on the existence of a well-defined bulk Hamiltonian. However, there are reasonable modifications of the CFT Hamiltonian $\mathcal{H}$ for which this will not be the case (see Fig.~\ref{fig-ambiguous}).  At the time $\tau_0$ on the boundary, let us add to $\mathcal{H}$ a source for the nonlocal CFT operator $\mathcal{O}$ dual to a local bulk operator at the origin, $\phi(\tau=0, \rho=0)$:
\begin{equation}
\mathcal{H} \to \mathcal{H} + J\delta(\tau-\tau_0) \mathcal{O} 
\end{equation}
We are working perturbatively, so the operator $\mathcal{O}$ is the one constructed using the above method and the unmodified Hamiltonian $\mathcal{H}$.  In the bulk, this source acts completely locally as a source for $\phi$ at the origin at $\tau=0$.  Note that this bulk point is spacelike related to the boundary slice $\tau_0$.  

Now let us compute the expectation value of the bulk operator $\phi$ at the origin, at some infinitesimal time $\epsilon$ after the source acts.  This operator can be constructed by the usual methods, but those methods require a knowledge of the bulk Hamiltonian in the region $S(\mathcal{S})$.  This Hamiltonian is ambiguous: from the viewpoint of the strip $\mathcal{S}$, the source does not act, since it acts in the CFT only at $\tau_0$.  Then the bulk evolution should be computed from the usual bulk Hamiltonian without source, and we obtain $\langle \phi(\epsilon,0)\rangle=0$.   From the viewpoint of the CFT at the time $\tau_0$, the bulk region near $(0,0)$ contains a source.  Then the bulk evolution should take this source into account and we obtain  $\langle \phi(\epsilon,0)\rangle\neq 0$.  Thus, there is no unique assignment of a bulk field value at $(\epsilon,0)$.  In this sense, $(\epsilon,0)$ should not be considered a bulk point dual to the strip.

\begin{figure}[h]
\centering
\includegraphics[width=3in]{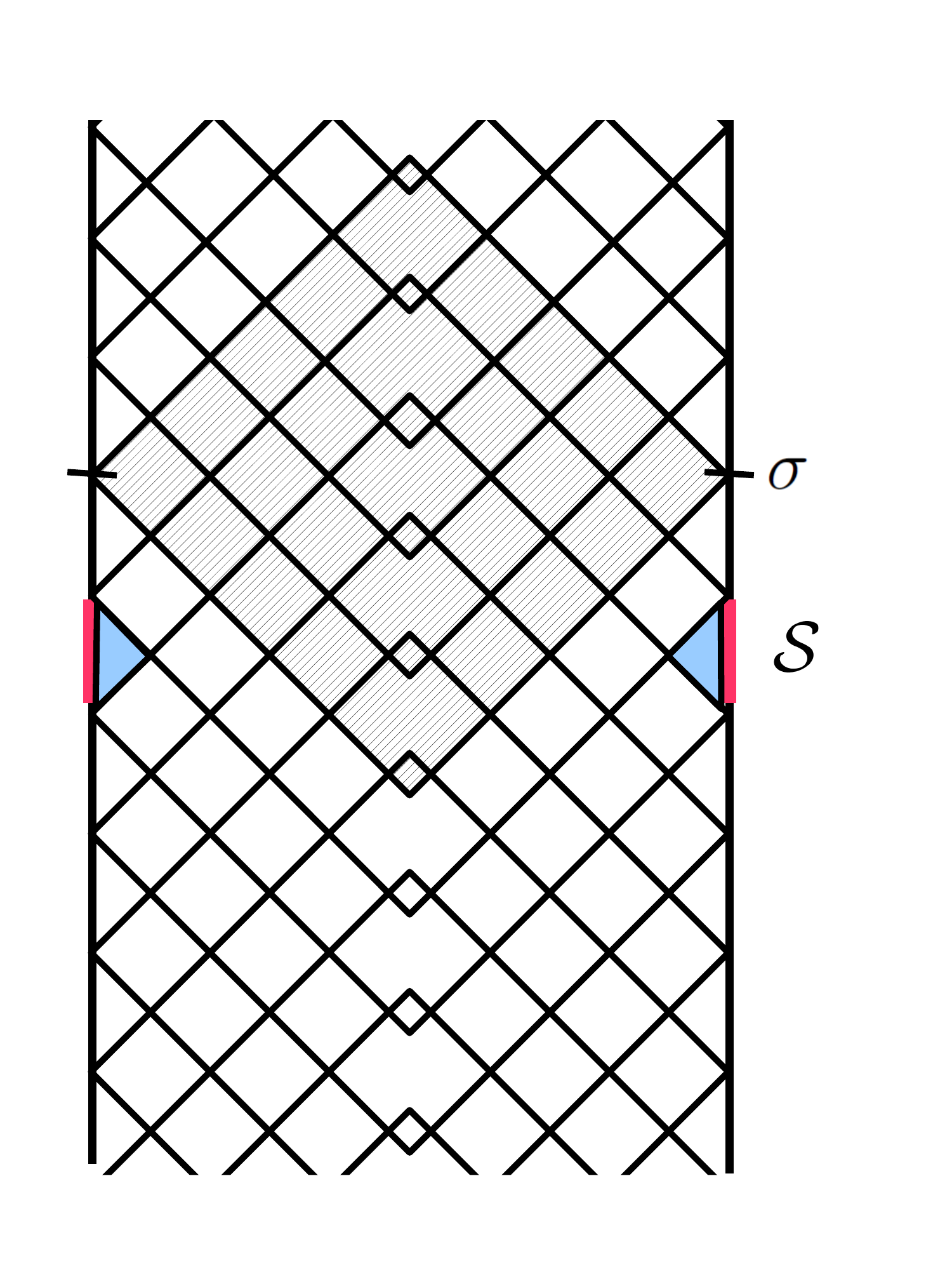}
\caption{The shaded region shows bulk points spacelike related to a global boundary Cauchy surface $\sigma$.  The union of all such sets over the collection of boundary Cauchy surfaces which do not intersect $\mathcal{S}$ has an ambiguous bulk interpretation when the boundary Hamiltonian is allowed to vary outside of $\mathcal{S}$. The unambiguous region, $\bar{H}(\mathcal{S})$, is the complement of this union. In this example, we see that $\bar{H}(\mathcal{S}) = C(\mathcal{S})$.}
\label{fig-ambiguous2}
\end{figure}

Let us now consider a general boundary region $b$ and construct a bulk region $\bar{H}$ such that the interpretation of what happens in $\bar{H}$ is unambiguous (Fig.~\ref{fig-ambiguous2}).  The ambiguities we discussed arise from inserting nonlocal CFT operators into the Hamiltonian on Cauchy surfaces of the boundary which do not intersect the region $b$. A modification of the CFT Hamiltonian on such a Cauchy surface $\sigma$ can lead to an ambiguous bulk Hamiltonian in the region $S(\sigma)$ spacelike-related to $\sigma$.  Thus we find that $\bar{H}$ is the compliment of the union of all $S(\sigma)$, where $\sigma$ is any Cauchy surface for the global boundary which lies in the complement of $b$. 

The region $\bar{H}$ is closely related to $C(b)$ and hence to $H(b)$. It is easy to see that $H=C\subset \bar{H}$. The study of a number of examples suggests that $H\neq \bar H$ if and only if an event horizon is present in the bulk. It would be nice to study $\bar H$ and its relation to $H$ further.  The discussion in the following section may be relevant.

\section{Quantum Gravity Behind Event Horizons}
\label{sec-bh}

In this section we discuss an issue that is somewhat orthogonal to the main subject of this paper: the degree to which the reconstruction of bulk regions, perturbatively in $1/N$, allows us to claim that AdS/CFT provides a full quantum gravity theory for regions behind event horizons, such as the interior of a black hole, or cosmological regions.  We will construct an experiment behind the horizon whose outcome is known but not captured by such methods.

The CFT provides a full quantum gravity theory for observers near the boundary.\footnote{Note that all observers that remain outside the black hole have the same causal diamond, consisting of the exterior of the black hole.  The covariant RG flow we described in Sec.~\ref{sec-RG} can be thought of as moving the observers deeper into the bulk, but note that the flow never enters the black hole.}  It completely settles the issue of whether the formation and evaporation of a black hole is a unitary process.  It is crucial for this argument that the time evolution is carried out on the boundary, where it is manifestly unitary; the duality is used only at early and late times in order to recover the in and out states in the bulk~\cite{Sus98, Pol99, Gid99, Fitz11}.

To what extent can we regard the CFT {\em also\/} as a full quantum gravity theory for an observer falling into the black hole?  Perhaps, by reordering the degrees of freedom, one could interpret the CFT as providing a nonperturbative definition of quantum gravity for the infalling observer?  This would require that the bulk dual region is ambiguous, at the nonperturbative level.  This may be the case, but the dictionary that would provide this definition is not known at the required level of precision.  

The black hole interior is clearly encoded in the CFT if one makes use of bulk equations of motion to evolve the infalling data back out of the black hole and to the boundary.  But in the same approximation, we can also generate a xeroxing paradox~\cite{SusTho93b}: at the semiclassical level, there is no manifest obstruction to evolving to global bulk slices that contain both the black hole interior and the Hawking cloud.  

Perhaps we should restrict the bulk evolution by hand to the causal patch of an infalling observer, and settle for this approximate description of the black hole interior?  The finiteness of entropy bounds inside black holes imply that there cannot be exact observables associated with the infalling observer at late times.  Thus, the description of the infalling observer should be less precise that that of the observer near the boundary, who has access to exact observables.  Perhaps the need to use bulk evolution is simply a reflection of this intrinsic limitation?

In fact, however, it is clear that the infalling observer requires a theory that goes beyond semiclassical bulk evolution.  This can be seen from the following thought-experiment.  Consider an infalling observer who performs scattering experiments inside a Schwarzschild-AdS black hole of radius $R$, after crossing the event horizon.  The scattering occurs at high enough energy to produce a (smaller) black hole of mass $m$, which then evaporates.  We are free to choose parameters so that the scattering effectively takes place in flat space:
\begin{equation}
R_{\rm AdS}\gg R\gg f(m) \gg 1~,
\end{equation}
where $f(m)$ is at least the evaporation timescale of the black hole ($m^d$ in $d$ spatial bulk dimensions).  For sufficiently large but finite $R/f(m)$, the infalling observer can confirm the unitarity of black hole formation and evaporation to any required precision.  But this conflicts with the result that would be obtained from the semiclassical analysis (Hawking's calculation).  

We conclude that the proper description of an infalling observer requires a quantum gravity theory beyond the semiclassical approximation.  The CFT on the boundary does not provide us with that theory, since its application to the infalling observer relies on semiclassical bulk evolution.  No other way of relating local operators inside the black hole to boundary operators is known; therefore, we cannot replace this bulk evolution by boundary evolution, as we did in the case of scattering experiments performed by a near-boundary observer.

Of course, in the limit used in the thought-experiment, one could imagine ``cutting out'' the spacetime region of size $f(m)$ that contains the scattering experiment.  One could embed this region in an asymptotically AdS spacetime with $R_{\rm AdS}\gg f(m)$ and use the flat space S-matrix, which can be computed using AdS/CFT~\cite{Sus98, Pol99, Gid99, Fitz11}.  In fact, this argument is what gives us confidence that the process is indeed unitary.  But how would this prescription generalize?  A theory of the infalling observer that relied on this type of cutting and pasting would not be applicable to the highly dynamical regions deeper inside the black hole, nor to the spacelike singularity, which cannot be so transplanted.

An exact version of this cut-and-paste process is available if the black hole is formed by sending in a spherical null shell from the boundary of AdS~\cite{Hub02}. By causality, the bulk region in the past of the shell is the same no matter whether we decide to send in the shell or not.  If we do send in the shell, then this region includes a portion $w$ of the black hole interior.  If we do not, then the same bulk region $w$ is dual to operators on the boundary, which can be evolved to nonlocal operators $W$ on a single global boundary time slice preceding the insertion of the shell.  This illustrates that cut-and-paste is well-defined precisely in the limit where it yields no information about regions that could {\em only\/} exist in the interior of black holes, such as singularities and a highly dynamical geometry.

The absence (so far) of an intrinsic bulk theory at the nonperturbative level appears to impose crucial limitations on our ability to describe black hole interiors and cosmological regions\footnote{A similar conclusion~\cite{Gar10} applies to the interior of other event horizons, such as an FRW universe~\cite{Mal10}.  In this case there are two natural choices of conformal frame on the boundary~\cite{BarRab11}: one in which the CFT is well-behaved, and another in which the coefficient of a relevant operator diverges in finite time.  It is tempting to interpret this violent behavior in terms of an infalling observer hitting the big crunch singularity behind the horizon, but it can be understood more simply as the arrival of the bulk domain wall on the boundary.} via AdS/CFT, beyond what follows from the approximate methods that were already available to generate bulk evolution.   (See, however, Refs.~\cite{KraOog03,FidHub04,FestLiu,HorAlb09} for interesting approaches to this problem.)

\acknowledgments We thank N.\ Arkani-Hamed, T.\ Banks, B.\ Freivogel, J.\ Kaplan, H.\ Liu, J.\ Maldacena, D.\ Marolf, J.\ Polchinski, E.\ Rabinovici, S.\ Shenker, A.\ Strominger, L.\ Susskind, and E.\ Witten for discussions. This work was supported by the Berkeley Center for Theoretical Physics, by the National Science Foundation (award numbers 0855653 and 0756174), by fqxi grant RFP3-1004, and by the US Department of Energy under Contract DE-AC02-05CH11231. The work of VR is supported by an NSF Graduate Fellowship. The work of SL is supported by a John A.\ McCone Postdoctoral Fellowship.

\appendix

\section{Light-sheets from the conformal boundary of AdS}\label{sec-normal}

In this appendix we show that the light-rays orthogonal to any spacelike slice of the conformal boundary of AdS have nonpositive expansion into the bulk in the physical metric, and thus generate light-sheets.  This is an important property of light-sheets in asymptotically AdS spacetimes. It guarantees that the maximum entropy of the bulk holographic domain $H$ is nonincreasing under the covariant RG flow described in Sec.~\ref{sec-RG}. 

In Poincar\'e coordinates, the AdS metric is:
\begin{equation} \label{eq:ads}
ds^2 = \frac{1}{z^2}(dz^2 + \eta_{\mu \nu} dx^{\mu} dx^{\nu})~,
\end{equation}
where $\eta_{\mu \nu}$ is the metric for $d$ dimensional Minkowski space. 
Consider the conformally rescaled metric
\begin{equation}
d\tilde{s}^2 = dz^2 + \eta_{\mu \nu} dx^{\mu} dx^{\nu}~,
\end{equation}
which is $d+1$ dimensional Minkowski space. Take some small region of the boundary and let $\tilde{\theta}$ be the expansion of some congruence of infinitesimally neighboring light rays in this space. That is
\begin{equation} \label{eq:theta}
\tilde{\theta} = \frac{d \log \mathcal{\tilde A}}{d \tilde{\lambda}}~,
\end{equation}
where $\tilde{\mathcal A}$ is the infintesimal area spanned by the light rays and $\tilde{\lambda}$ is the affine parameter. In order for the null geodesics to remain affinely parameterized after the conformal transformation, the affine parameter must transform as (Appendix D of ~\cite{Wald}) 
\begin{equation} \label{eq:lambda}
\frac{d \tilde{\lambda}}{d \lambda} = c z^2~,
\end{equation}
where $c$ is a constant and $z^2$ is the conformal factor. The area will transform as
\begin{equation} \label{eq:A}
\mathcal{A} = \frac{\mathcal{\tilde{A}}}{z^{d-1}}.
\end{equation}

Using (\ref{eq:theta}), (\ref{eq:lambda}), and (\ref{eq:A}) we find that the expansion in the AdS spacetime (\ref{eq:ads}) is
\begin{equation}
\theta =   c z^2\tilde{\theta} - c(d-1)z\tilde{k}^z~,
\end{equation}
where 
\[
\tilde{k}^z = \frac{d z}{d \tilde{\lambda}}~.
\]
Since both $\tilde{\theta}$ and $\tilde{k}^z$ are defined in $d+1$ dimensional Minkowski space using a congruenece of null rays orthogonal to a spacelike (and hence nowhere null) foliation, they are finite. Thus, at the boundary ($z=0$), we have $\theta = 0$.

\bibliographystyle{utcaps}
\bibliography{all}

\end{document}